\documentclass[aps,epsfig,nofootinbib]{revtex4}
\usepackage{graphicx}
\usepackage{amsmath}
\usepackage{amssymb}
\usepackage{epsfig}
\usepackage[colorlinks,linkcolor=blue,citecolor=blue,urlcolor=blue]{hyperref}

\begin{document}

\title{Revisiting pre-inflationary universe of family of $\alpha-$attractor in loop quantum cosmology}
\author{M. Shahalam$^{1,2}$ \footnote{E-mail address: shahalam@zjut.edu.cn}}
\author{Mudhahir  Al  Ajmi$^3$ \footnote{E-mail address: mudhahir@squ.edu.om}}
\author{R. Myrzakulov$^2$ \footnote{ E-mail address: rmyrzakulov@gmail.com}}
\author{Anzhong Wang$^{4}$ \footnote{E-mail address: Anzhong$\_$Wang@baylor.edu}}
\affiliation{$^{1}$Institute for Theoretical Physics $\&$ Cosmology,
Zhejiang University of Technology, Hangzhou, 310023, China\\
$^2$Eurasian International Center for Theoretical Physics, Department of General and Theoretical Physics, L. N. Gumilyov Eurasian National
University, Nur Sultan, 010008, Kazakhstan\\
$^3$Department  of  Physics,  College  of  Science,  Sultan  Qaboos  University, P.O.  Box  36,  Al-Khodh  123,  Muscat,  Sultanate  of  Oman\\
$^4$GCAP-CASPER, Department of Physics, Baylor University, Waco, TX, 76798-7316, USA }

\date{\today}

\begin{abstract}
In this work, we revisit the dynamics of pre-inflationary universe with a family of $\alpha-$attractor potentials, in the framework of loop quantum cosmology, in which the big bang singularity is generically resolved purely with quantum geometric effects, and replaced by a quantum bounce. At the bounce, the background evolution is divided into two distinct classes, the first is dominated by the kinetic energy of the inflaton field and the second by the potential energy. In both classes, we find the physically viable initial conditions numerically that provide not only the slow-roll inflation, but also sufficient e-folds to be compatible with observations. In the entire
range of kinetic energy dominated initial conditions  (except some subsets of Models 2 and 4), the background evolution prior to reheating is always split into three different 
phases: bouncing, transition and slow-roll inflation. In the bouncing phase, the numerical evolution of the scale factor is independent not only of the initial data, but also the inflationary potentials, as long as it is dominated by the kinetic energy, and can be well approximated by an analytical solution, whereas in the potential energy dominated case, such approximated results do not exist. Moreover, we study the phase space analysis for a class of $\alpha-$attractor potentials, and discuss the phase space trajectories for physically viable initial conditions of the inflaton field. 
\end{abstract}


\keywords {Inflation, Loop quantum cosmology}

\maketitle
\flushbottom

\section{Introduction}
\label{sec:intro}
The cosmic inflation has emerged as a successful paradigm to resolve  various issues in the standard model of cosmology, including  the horizon and flatness problems. Inflation can  explain the origin of inhomogeneities observed in cosmic microwave background and the structure formation of the universe \cite{guth1981}. A large number of inflationary models have been proposed in the literature such as conformal attractor \cite{conformal}, $\alpha-$attractor \cite{alpha,alpha1,alpha2,alpha3,alpha4}, Starobinsky and the chaotic inflation \cite{staro1980,staro1,staro2,staro3,staro4,GL}. The cosmological predictions of these models are very similar but not identical as the main difference is in the shape of the potentials. These models are in good agreements with the present observational data. In the case of a single field inflation, Starobinsky and $\alpha-$attractor potentials are fully consistent with the Planck 2018 data, whereas the quadratic potential is  ruled out \cite{Planck2018}. In this paper, we shall revisit the dynamics of the pre-inflationary universe with the class of $\alpha-$attractor potentials in the framework of loop quantum cosmology (LQC), and explore whether the slow-roll inflation is achieved or not followed by the initial quantum bounce. Recently, the similar results for the $\alpha-$attractor that contains $T$ and $E$ models have been studied in \cite{alamPRD2018}.

All inflationary models that are based on general relativity (GR) suffer from the initial and inevitable singularity \cite{borde1994,borde2003}. Therefore, it is difficult to know how and when to impose the initial conditions. In addition, the inflationary universe should have at least 60 $e$-folds to be consistent with observations. However, more than 70 $e$-folds can be found in a large class of inflationary models in which the size of present universe is smaller than the Planck at the beginning of inflation \cite{martin2014}. As a result, the semi-classical treatments are questionable in these models. This is known as the trans-Planckian problem \cite{martin2001,berger2013}. 

The above issues can be addressed in the framework of LQC, which provides a feasible explanation of inflation and pre-inflationary dynamics simultaneously. It is remarkable to note that in such a framework the big bang singularity is replaced by a non-singular quantum bounce \cite{agullo2013a,agullo2013b,agullo2015,ashtekar2011,ashtekar2015,barrau2016}. Furthermore, universe that onsets at the quantum bounce usually enters in the slow-roll inflation \cite{ashtekar2010,psingh2006,zhang2007,chen2015,bolliet2015,schander2016,bolliet2016,Bonga2016,Mielczareka}. For the pre-inflationary universe, in the framework of LQC, two main approaches are discussed in the literature, the dressed metric \cite{agullo2013b,metrica,metricb,metricc} and the deformed algebra \cite{algebraa,algebrab,algebrac,algebrad,algebrae,algebraf}. For the background evolution, both approaches provide the same set of evolution equations but their perturbations are distinct \cite{bolliet2016}. The corresponding non-Gaussianities were investigated in \cite{agullo15,ABS17,ZWKCS18}. 

In this work, we consider a family of $\alpha-$attractor potentials, and are mainly interested in the background evolution of the universe. Therefore, the results to be obtained in this paper will be valid to both approaches. Specially, we shall exhibit that, for the kinetic energy dominated (KED) initial conditions, the evolution of the universe before reheating can be divided into three different phases: {\em bouncing, transition and slow-roll inflation},  while this is not possible in the potential energy dominated (PED) case \cite{alamPRD2018,alam2017,Tao2017a,Tao2017b}. The analytical evolution of the background and linear perturbations during these phases have been discussed in \cite{Tao2017a,Tao2017b}. Moreover, many authors have studied  various inflationary models in LQC, GR, string-inspired models and Bianchi I universe \cite{yang2009,DL17,adlp,lsw2018a,lsw2018b,agullo18,thiemann,HISY,BG15,sahni18,SW08,killian,nozari}, 
\cite{BaoFei2019a,BaoFei2019b,wu2018,ma2019,anshu2019,Bea2018,sharma2018,ye2018},
and important results were discussed.
 
The rest of the paper is organized as follows. In Sec. \ref{sec:alphamod}, the family of $\alpha-$attractor potentials is briefly discussed with four new models. In sec. \ref{sec:EOM}, we study the background equations of the Friedmann-Lemaitre-Robertson-Walker (FLRW) universe in the framework of LQC. The Subsections \ref{subsec:mod1}, \ref{subsec:mod2}, \ref{subsec:mod3} and \ref{subsec:mod4} are devoted to the detailed analysis of the background evolution with $\dot{\phi_B}>0$, and also for the kinetic energy (KE) and potential energy (PE) dominated initial conditions at the quantum bounce. The phase portraits are displayed in Sec. \ref{sec:port}. Our main results are summarized  in Sec. \ref{sec:conc}.

\section{A family of $\alpha-$models}
\label{sec:alphamod}

Following \cite{kalloshPRL15,linder15,alam2018}, the Lagrangian density of the
$\alpha-$attractor models with non-canonical kinetic term and a potential is given as
\begin{equation}
\mathcal{L}=\sqrt{-g}\left[ \frac{1}{2} M_{Pl}^2R-\frac{\alpha }{\left( 1-\frac{\varphi ^{2}}{6}%
\right) ^{2}}\frac{\left( \partial \varphi \right) ^{2}}{2}-\alpha
f^{2}\left( \frac{\varphi }{\sqrt{6}}\right) \right]
\label{eq:lag}
\end{equation}
where $M_{Pl}=m_{Pl}/\sqrt{8 \pi}$ denotes the reduced Planck mass, $\alpha
f^{2}$ represents the potential function and  $\alpha$ is a parameter. The non-canonical kinetic term in Eq. (\ref{eq:lag}) can be made canonical through the field redefinition $\phi =\sqrt{%
6\alpha }\tanh ^{-1}\left( \frac{\varphi }{\sqrt{6}}\right)$. Therefore, the potential  is given by
\begin{equation}
V\left( \phi \right) =\alpha f^{2}\left( \tanh \left( 
\frac{\phi }{\sqrt{6\alpha }}\right) \right).
\label{eq:vf}
\end{equation}
Two functional forms of $f$ have been extensively used in the literature,
\begin{eqnarray}
\label{eq:modT}
f(x)&=&c x \\
f(x)&=&c \frac{x}{1+x}
\label{eq:modE}
\end{eqnarray}
where $x =\tanh \left( \frac{\phi }{\sqrt{6 \alpha}}\right)$, and $c$ is a constant that scales the amplitude of the potential. Eq. (\ref{eq:modT}) is known as $T$ model \cite{alpha,alpha2,alpha3}, and reduces to the Goncharov and Linde model for $\alpha=1/9$ \cite{GL}. Eq. (\ref{eq:modE}) is  the so-called $E$ model and reduces to   Starobinsky's model for $\alpha=1$ \cite{alpha1,staro1980}. The pre-inflationary universe and phase space analysis for $T$ and $E$ models in context of LQC have been examined in \cite{alamPRD2018}.

In this work, we shall choose the following functional forms of $f$, and investigate the pre-inflationary dynamics of the inflaton field in the framework of LQC. We shall examine whether these forms can lead to the desired slow-roll inflation or not, followed by the quantum bounce. These functional forms are
\begin{eqnarray}
\label{eq:funcform1}
f(x)&=&c \frac{1}{x} \\
\label{eq:funcform2}
f(x)&=&c \frac{1}{1+x} \\
\label{eq:funcform3}
f(x)&=&c \frac{1}{\sqrt{1-x^2}} \\
\label{eq:funcform4}
f(x)&=&c \frac{x^2}{\sqrt{1-x^2}}
\end{eqnarray}
The right hand side of equations (\ref{eq:funcform1}), (\ref{eq:funcform2}), (\ref{eq:funcform3}) and (\ref{eq:funcform4}) blows up at $x=0, -1, 1$ and 1,  respectively. Furthermore, equation (\ref{eq:funcform4}) vanishes at $x=0$.

\begin{figure*}[tbp]
\begin{center}
\begin{tabular}{cc}
{\includegraphics[width=2.2in,height=1.7in,angle=0]{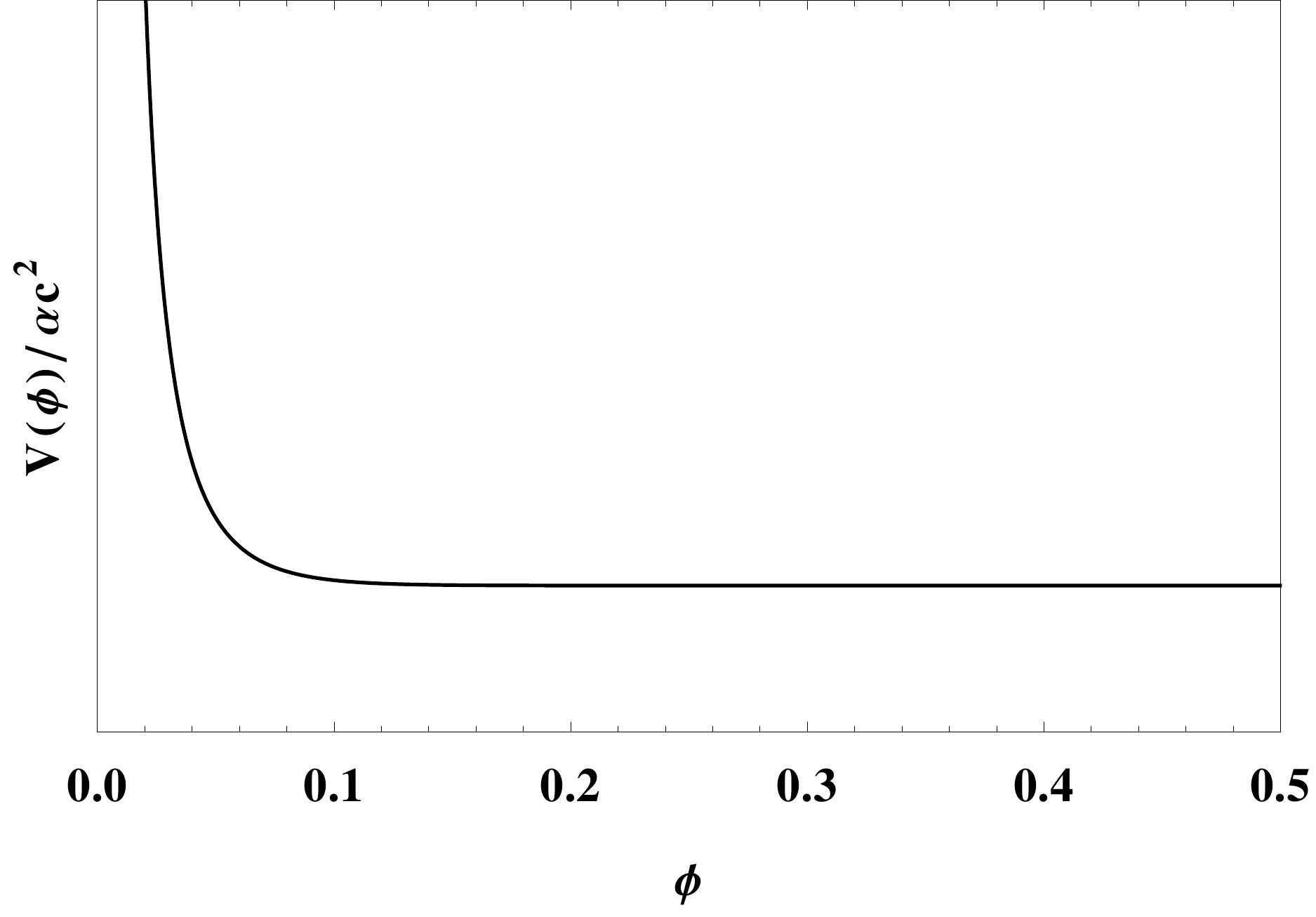}} &
{\includegraphics[width=2.2in,height=1.7in,angle=0]{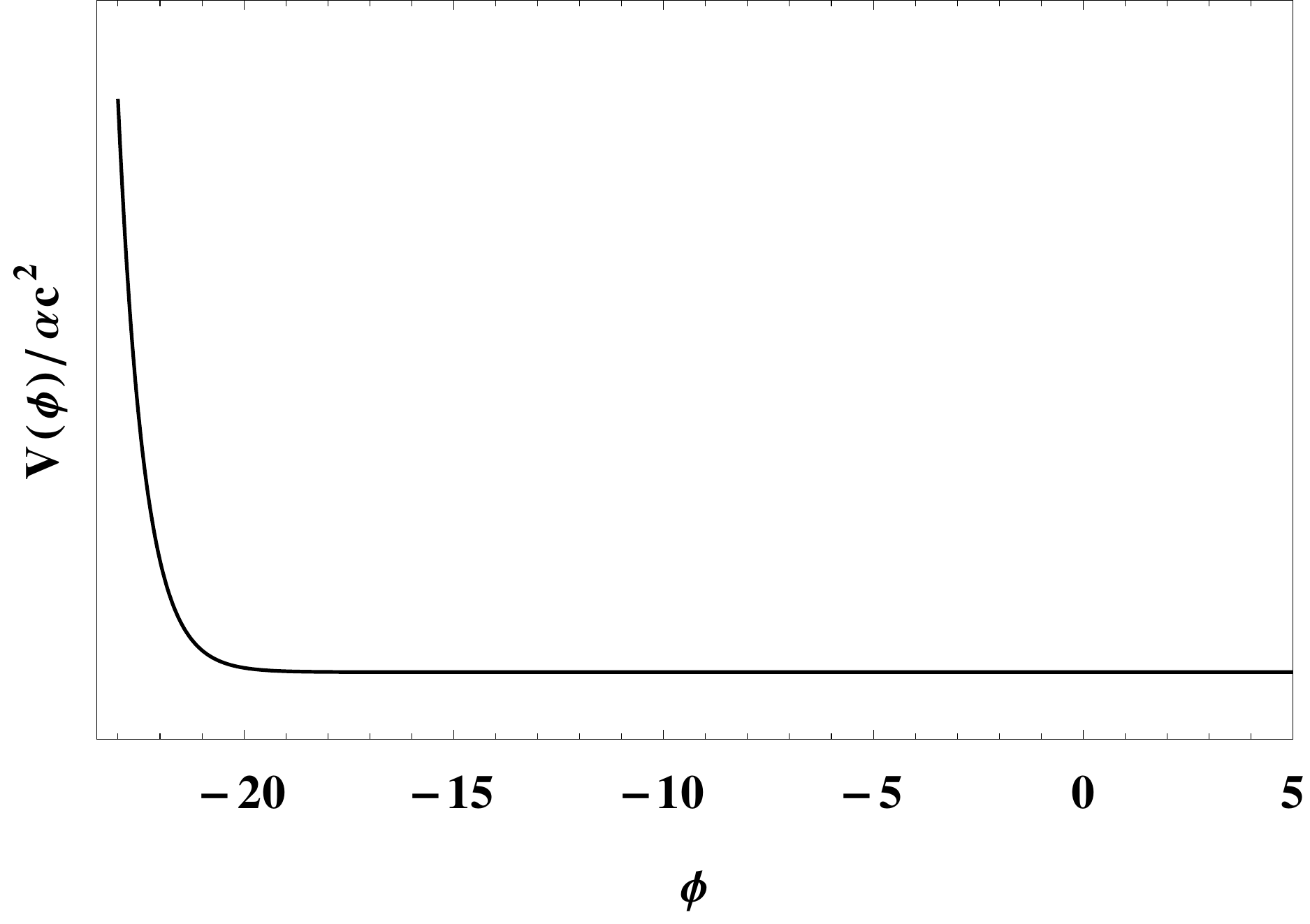}} 
\\
{\includegraphics[width=2.2in,height=1.7in,angle=0]{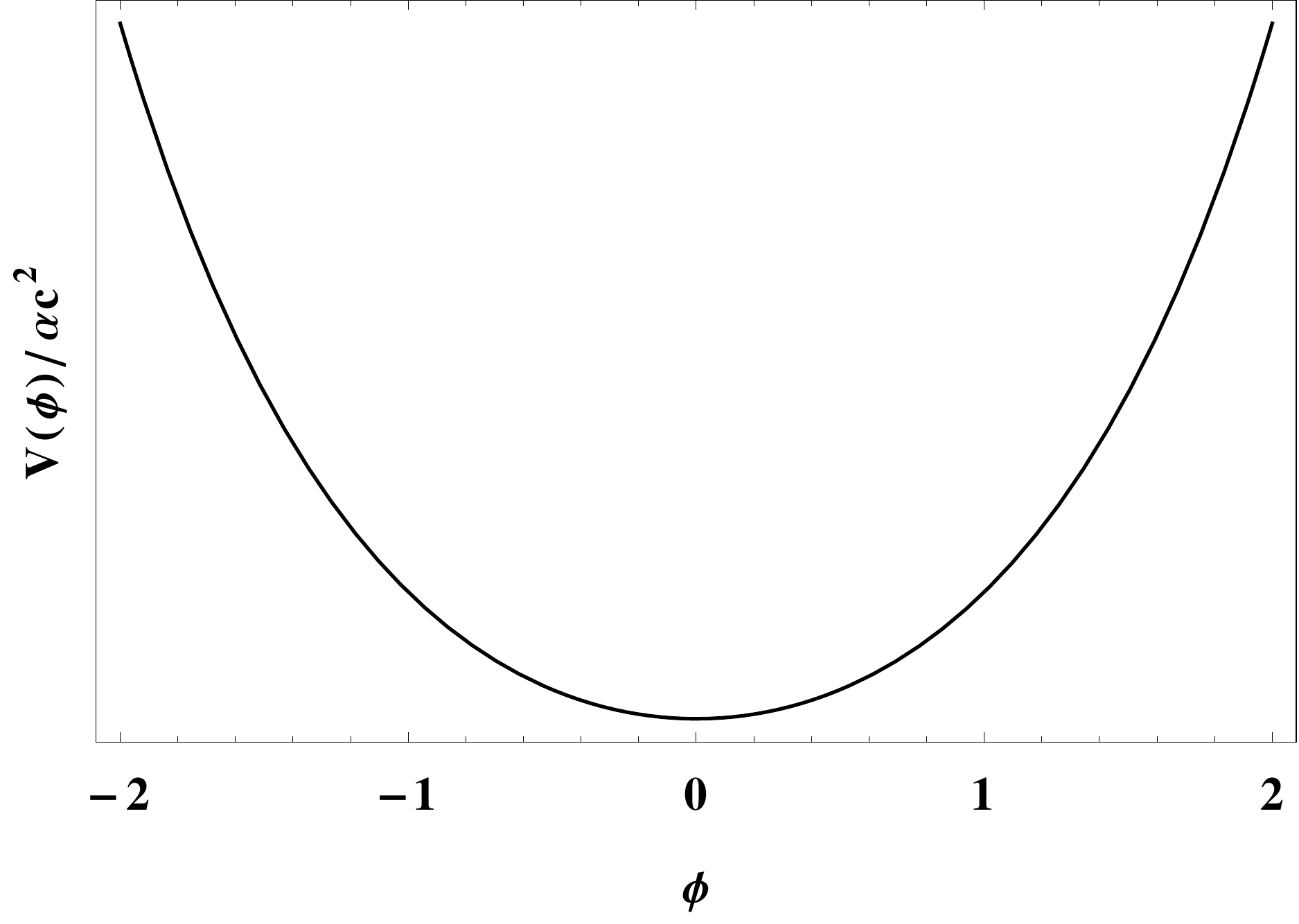}} & 
{\includegraphics[width=2.2in,height=1.7in,angle=0]{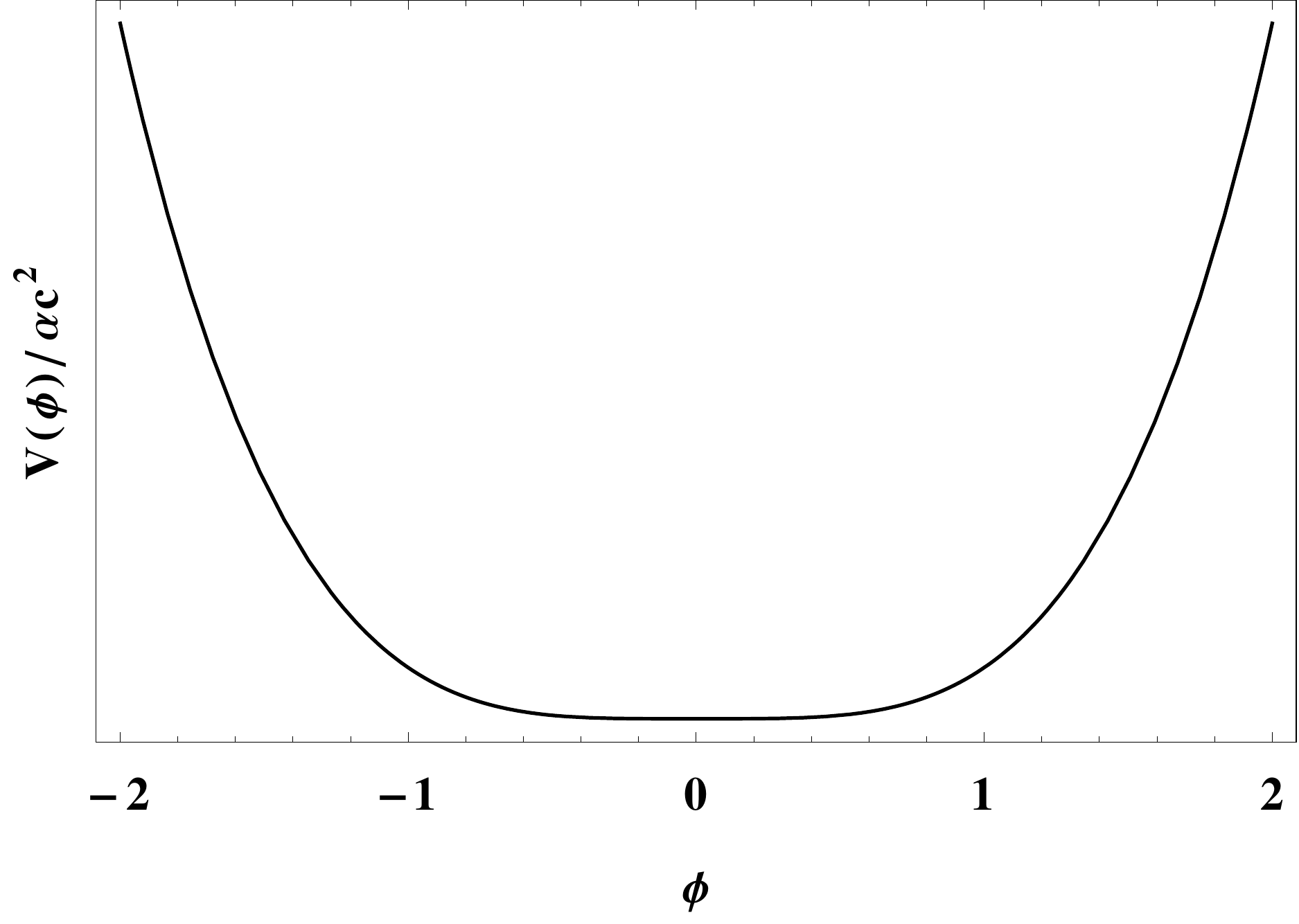}} 
\end{tabular}
\end{center}
\caption{This figure is schematically displayed for the models under consideration. Upper left and right panels exhibit the evolution of potentials (\ref{eq:pot1}) and (\ref{eq:pot2}). Both potentials blow up at $\phi=0$ and $\phi=- \infty$, respectively, while monotonically decline to constant behavior as $\phi \rightarrow \infty$. Lower left and right panels correspond to the evolution of potentials (\ref{eq:pot3}) and (\ref{eq:pot4}). Both potentials are symmetric with respect to $\phi=0$, and show oscillating behavior around the origin. For $\phi \rightarrow 0$, potentials (\ref{eq:pot3}) and (\ref{eq:pot4}) are bounded below by unity ($V(\phi) \geq 1$) and zero ($V(\phi) \geq 0$), receptively whereas for $\phi \rightarrow \pm \infty$ they are unbounded. In LQC, the maximum energy density is $\rho_c$ that constraints the value of the field at the bounce. More details are given in the subsections \ref{subsec:mod1}, \ref{subsec:mod2}, \ref{subsec:mod3} and \ref{subsec:mod4}. }
\label{fig:pot}
\end{figure*}

The potentials corresponding to equations (\ref{eq:funcform1}), (\ref{eq:funcform2}), (\ref{eq:funcform3}) and (\ref{eq:funcform4}) are 
\begin{eqnarray}
\label{eq:pot1}
V(\phi) &=& \alpha c^2~ \left[ \coth \left( \frac{\phi }{\sqrt{6\alpha }}\right)\right]^2\\
\label{eq:pot2}
V(\phi) &=& \frac{\alpha c^2}{4}~ \left[ 1+ \text{exp}\left(-\sqrt{\frac{2}{3\alpha}}\phi \right) \right]^2\\
\label{eq:pot3}
V(\phi) &=& \alpha c^2~ \left[ \cosh \left( \frac{\phi }{\sqrt{6\alpha }}\right)\right]^2\\
\label{eq:pot4}
V(\phi) &=& \alpha c^2~ \left[ \tanh \left( \frac{\phi }{\sqrt{6\alpha }}\right)\right]^4 \left[ \cosh \left( \frac{\phi }{\sqrt{6\alpha }}\right)\right]^2
\end{eqnarray}
Hereafter, we shall refer equations (\ref{eq:pot1}), (\ref{eq:pot2}), (\ref{eq:pot3}) and (\ref{eq:pot4}) to as models 1, 2, 3 and 4,  respectively. The evolutions of these models are shown in Fig. \ref{fig:pot}. Models 1 and 2 blow up at $\phi=0$ and $\phi=- \infty$, respectively. Both models monotonically decline to a constant value as $\phi \rightarrow \infty$. Models 3 and 4 show oscillating behaviors as the field approaches to the origin ($\phi=0$), and are symmetric with respect to the point $\phi=0$. In the context of dark energy, theses models have been studied in \cite{varun2018}.

\section{Background equations and numerical evolution}
\label{sec:EOM}

In LQC, the modified Friedmann equation in a spatially flat FLRW universe, and the Klein-Gordon equation with a single scalar field are given, respectively, by \cite{ashtekar2006}
\begin{eqnarray}
H^2=\frac{8 \pi}{3 m_{Pl}^2}~\rho \Big{(}1-\frac{\rho}{\rho_c}\Big{)}, 
\label{eq:Hub}
\end{eqnarray}
\begin{eqnarray}
\ddot{\phi}+3H \dot{\phi}+ \frac{dV(\phi)}{d\phi}=0,
\label{eq:ddphi}
\end{eqnarray}
where $H=\dot{a}/a$ denotes the Hubble parameter, $\rho=\dot{\phi}^2/2+V(\phi)$ is the energy density of the scalar field, and $\rho_c \simeq 0.41 m_{pl}^4$ \cite{Meissne,Domagala} represents the critical energy density. From equation (\ref{eq:Hub}) one can see that $H=0$ at $\rho=\rho_c$. This implies that the quantum bounce occurs at $\rho=\rho_c$. 

The background evolution with a bouncing phase is of great interest, and one of the main tasks  is to show the existence of a desired slow-roll inflation with certain  initial conditions at the quantum bounce \cite{psingh2006,Mielczarek,zhang2007,chen2015,alam2017,Tao2017a,Tao2017b,ashtekar2011}. To this effect, we shall study ``bounce and slow-roll inflation'' with a family of $\alpha-$attractor models.

We  solve  Eqs.(\ref{eq:Hub}) and (\ref{eq:ddphi}) numerically with the initial conditions of $a(t)$, $\phi(t)$ and $\dot{\phi}(t)$ at   the quantum bounce,  at which we have
\begin{eqnarray}
&& \rho = \rho_c = \frac{1}{2}\dot{\phi}^2(t_B)+V(\phi(t_B)), \nonumber\\
&& \dot{a}(t_B)= 0, 
\label{eq:bounce}
\end{eqnarray}
where $t_B$ denotes the moment   at which the bounce occurs. From (\ref{eq:bounce}), we find
\begin{eqnarray}
\dot{\phi}(t_B) &=& \pm \sqrt{2 \Big{(} \rho_c - V(\phi(t_B)) \Big{)}}.
\label{eq:bounce2}
\end{eqnarray}
Without loss of the generality, one can take
\begin{eqnarray}
a(t_B) &=& 1.
\label{eq:bounce3}
\end{eqnarray}
From  Eq.(\ref{eq:bounce2}), one can see that for a given potential, the initial conditions will be described by $\phi_B$ only. Later, we shall find two cases: (a) positive inflaton velocity (PIV):~~ $\dot{\phi}_B > 0$;  and (b) negative inflaton velocity (NIV): ~$\dot{\phi}_B < 0$. In this paper, we shall focus only PIV. However, one can easily carry out a similar analysis for the NIV case. Hereafter, we shall denote $\phi(t_B)$ and $\dot{\phi}(t_B)$ by $\phi_B$ and $\dot{\phi}_B$, respectively. 

Finally, we define the following quantities that will be used in this paper \cite{alam2017,Tao2017a,Tao2017b}.

(1) The equation of state (EoS) $w(\phi)$ is defined as
\begin{eqnarray}
w(\phi) = \frac{\dot{\phi}^2/2-V(\phi)}{\dot{\phi}^2/2+V(\phi)}.
\label{eq:w}
\end{eqnarray}
In the slow-roll regime, we have $w(\phi)\simeq-1$.

To differentiate the KE and PE dominated initial conditions  at the bounce, we define the quantity $w^B$ as
\begin{equation}
w^B \equiv  w(\phi) \Big{\vert}_{\phi=\phi_B}
= \begin{cases}  > 0, \qquad \text{KE} > \text{PE}, \\
  = 0, \qquad \text{KE}=\text{PE}, \\
 < 0, \qquad \text{KE} < \text{PE}. \end{cases}
\label{eq:wb}
\end{equation}

(2) The slow-roll parameter $\epsilon_H$ is defined as
\begin{eqnarray}
\epsilon_H = - \frac{\dot{H}}{H^2}.
\label{eq:epsilon}
\end{eqnarray}
In the slow-roll region, we have $\epsilon_H \ll 1$.

(3) The number of $e$-folds $N_{inf}$ during the slow-roll inflation is expressed as
\begin{eqnarray}
N_{inf} = ln \Big{(} \frac{a_{end}}{a_i} \Big{)} =  \int_{t_i}^{t_{end}} H(t) dt \nonumber \\
 = \int_{\phi_i}^{\phi_{end}} \frac{H}{\dot{\phi}} d\phi \simeq \int_{\phi_{end}}^{\phi_i} \frac{V}{V_{\phi}} d\phi, 
\label{eq:Ninf}
\end{eqnarray}
where $a_i$ ($a_{end}$) exhibits the scale factor when the inflation  onsets  (ends),  that is $\ddot{a}(t_i) \gtrsim 0$ and  $w(\phi_{end})=-1/3$.

(4) The analytical expression  of the scale factor $a(t)$ during the bouncing regime can be expressed as \cite{alam2017,Tao2017a,Tao2017b}
\begin{eqnarray}
a(t) &=& a_B \left( 1+ \delta \frac{t^2}{t_{Pl}^2} \right)^{1/6},
\label{eq:a}
\end{eqnarray}
where $a_B=a(t_B)$, $\delta = {24 \pi \rho_c}/{m_{Pl}^{4}}$ is a dimensionless parameter, and $t_{Pl}$  represents the Planck time.

In the following subsections, we shall study the class of $\alpha-$attractor models for $\dot\phi_B > 0$ (PIV), and see whether following the bounce a desired slow-roll inflation generically exists or not.

\subsection{Model 1}
\label{subsec:mod1}

\begin{figure}[tbp]
\begin{center}
\begin{tabular}{ccc}
{\includegraphics[width=1.9in,height=1.65in,angle=0]{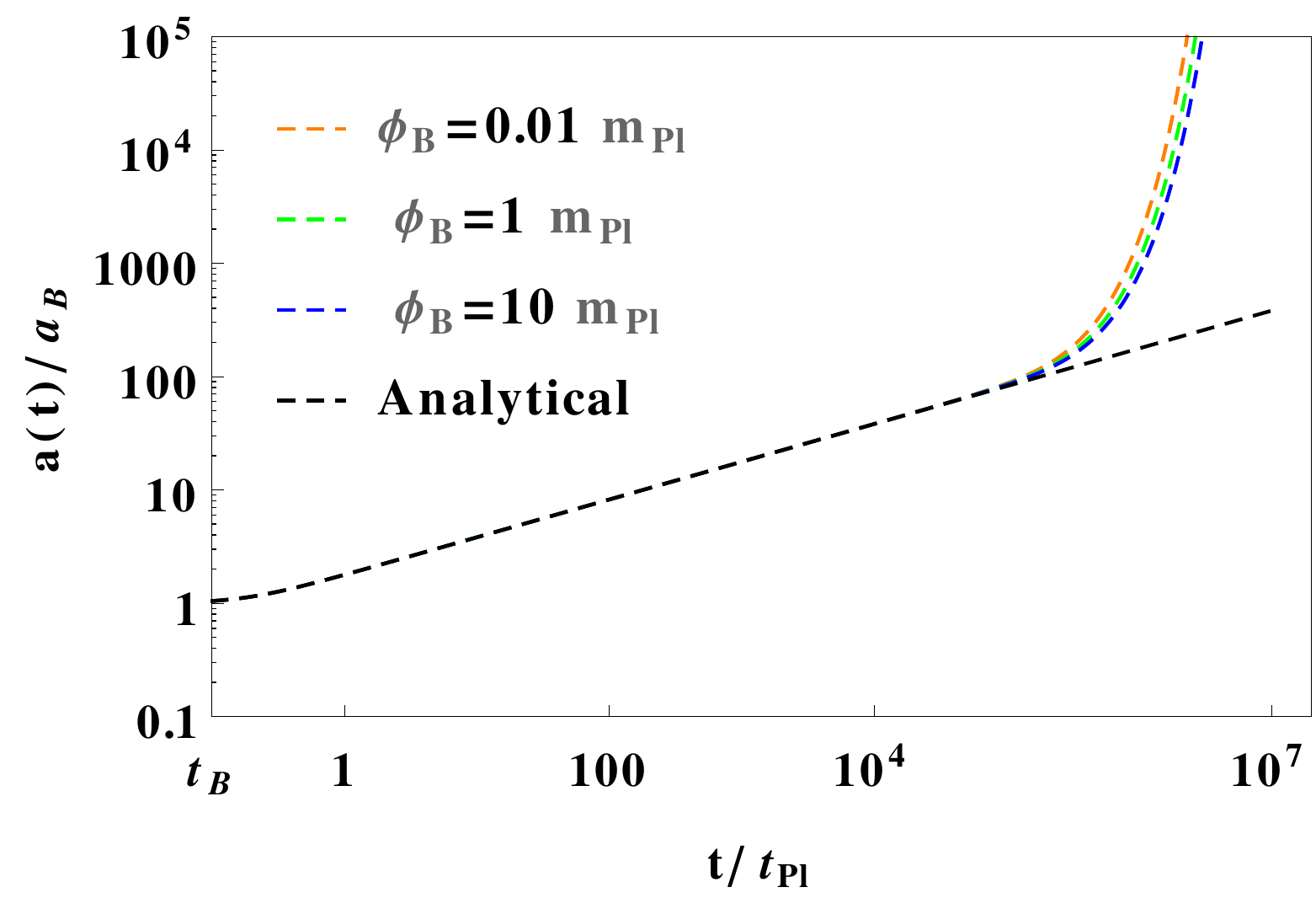}} &
{\includegraphics[width=1.9in,height=1.6in,angle=0]{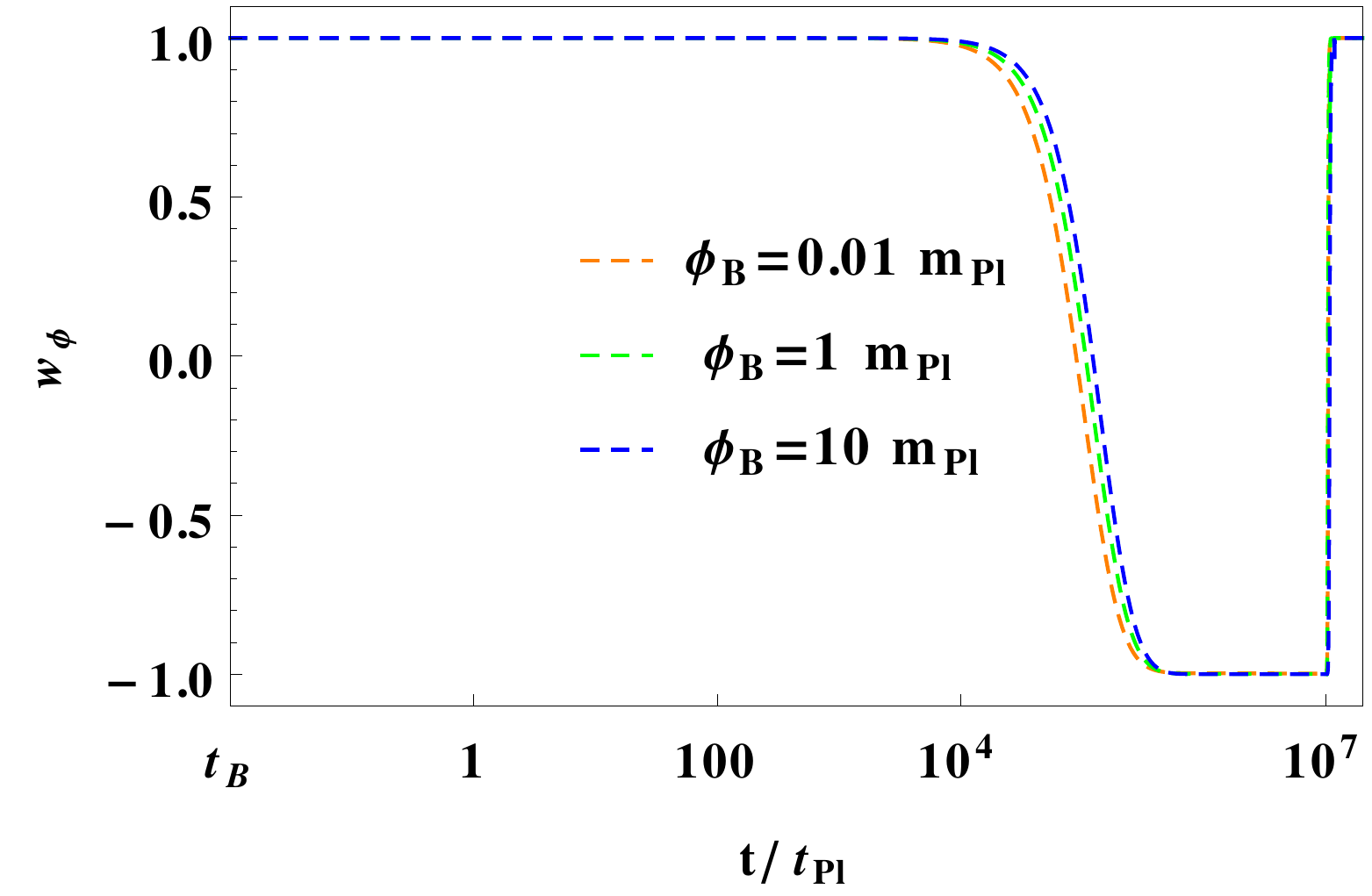}} &
{\includegraphics[width=1.9in,height=1.6in,angle=0]{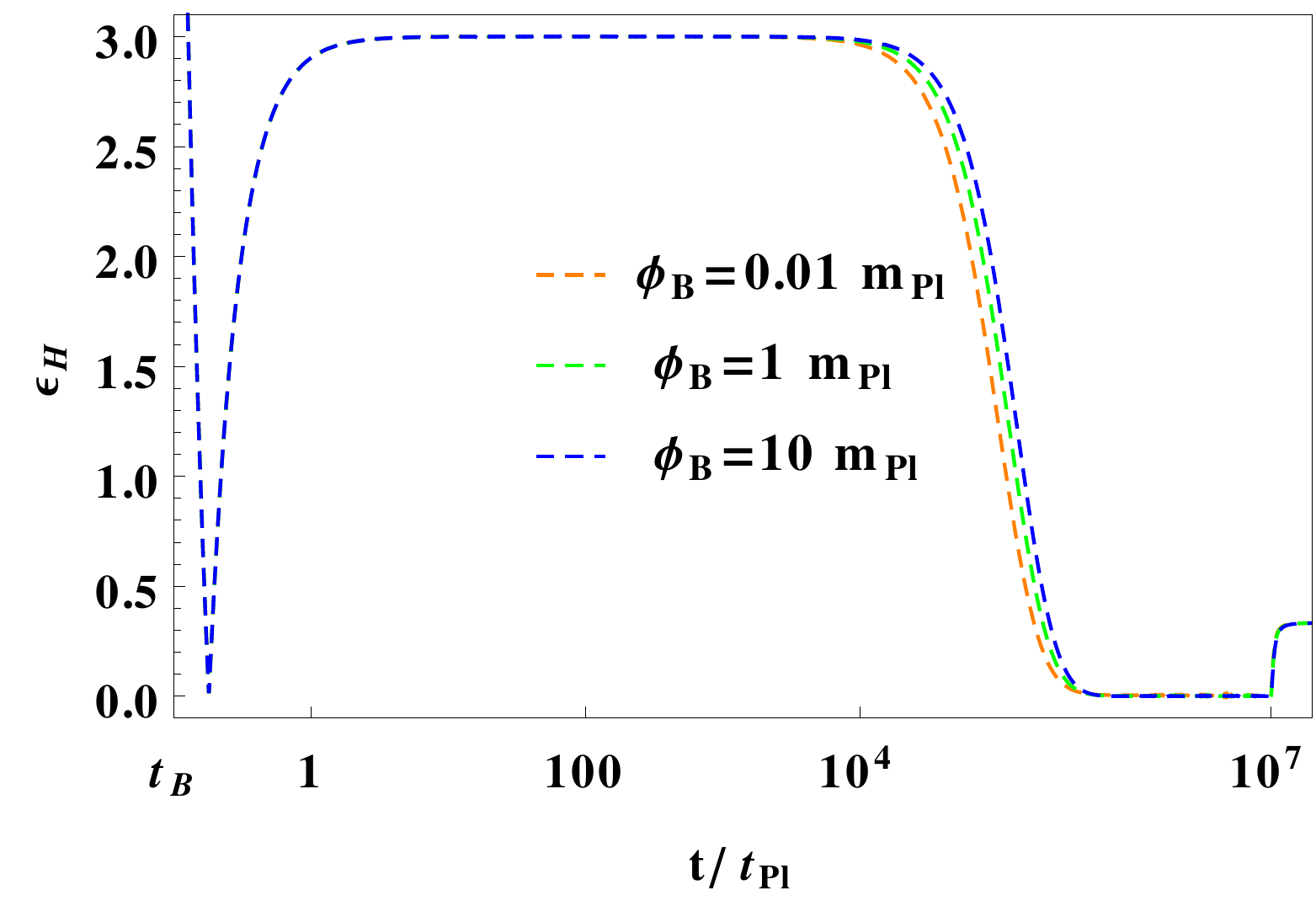}}
\end{tabular}
\end{center}
\caption{This figure represents the numerical results for model 1 [Eq.(\ref{eq:pot1})] with $\dot{\phi_B}>0$. The evolution of $a(t)$, $w(\phi)$ and $\epsilon_H$ is shown for the same set of the KED initial conditions of $\phi_B$ with $\alpha = 1 m_{Pl}^2$,  $c = 8.343 \times 10^{-7} m_{Pl}$ and $m_{Pl}=1$. The PED initial conditions are not possible to be imposed in the entire range of $\phi_B$. The analytical solution of the scale factor $a(t)$ [Eq.(\ref{eq:a})] is also exhibited in order to compare it with the numerical results. }
\label{fig:mod1}
\end{figure}

Let us first study some features of model 1 [Eq.(\ref{eq:pot1})]. The evolution of the potential (\ref{eq:pot1}) vs the scalar field is shown in the upper left panel of Fig. \ref{fig:pot}. This potential becomes asymptotically flat for the large field limit ($\phi \rightarrow \infty$), and blows up at the origin ($\phi =0$). In LQC, the maximum energy density is $\rho_c$ that constraints the value of $\phi_B$ as $(\phi_{min}, \infty)$,
where
\begin{eqnarray}
\phi_{min} &\simeq & \sqrt{6 \alpha}~ \text{arccoth} \left( \sqrt{\frac{\rho_c}{\alpha c^2}} \right).
\label{eq:mod1phimin}
\end{eqnarray}
To find the values of $\alpha$ and $c$ that are consistent with the Planck 2018 data for an inflationary universe \cite{Planck2018}, we follow the prescription provided in Appendix A. In particular, choosing $H_* = 2.0\times 10^{-5} M_{Pl}$, we can find $\phi_*$ from Eq.(\ref{eq:HubSR}) for the given potential in this model.  Then, setting $\epsilon_V = 1$ in Eq.(\ref{eq:ev}) we find $\phi_{end}$. With such obtained $\phi_*$ and $\phi_{end}$, we can find $(\alpha, c)$ from Eq.(\ref{eq:Ninf2}) by setting $N_{inf} = 60$. In doing so, we find various sets of $(\alpha, c)$,  which are all consistent with the Planck 2018 data. All of these cases give  similar conclusions. So, in the following we shall consider only one representative case, which is given by
\begin{eqnarray}
\alpha &=& 1 m_{Pl}^2, \qquad\qquad c = 8.343 \times 10^{-7} m_{Pl}.
\label{eq:mod1alphac}
\end{eqnarray}
Then, we numerically solve Eqs. (\ref{eq:Hub}) and (\ref{eq:ddphi}) with PIV ($\dot{\phi}_B>0$) for model 1. The results for a set of KED initial conditions with $\alpha = 1 m_{Pl}^2$ and  $c = 8.343 \times 10^{-7} m_{Pl}$ are shown in Fig. \ref{fig:mod1}, where the scale factor $a(t)$, EoS $w(\phi)$, and  slow-roll parameter $\epsilon_H$ are exhibited for the same set of $\phi_B$. The initial values of inflaton field at the bounce are governed by the KED conditions with the entire range of $\phi_B$, while the PED initial conditions are not possible at all in the whole range. Similar results were discussed for $T-model$ in Ref. \cite{alamPRD2018}.  

From the middle panel of Fig. \ref{fig:mod1}, one can clearly see that the evolution of the universe before reheating can be divided into three distinct phases: bouncing, transition and slow-roll inflation. In the bouncing phase, KE dominates,  and $w(\phi) \simeq +1$. During the transition region, $w(\phi)$ decreases rapidly  from $+1$ $(t/t_{Pl} \simeq 10^4)$ to $-1$ $(t/t_{Pl} \simeq 10^5)$. This transition phase is very short in comparison with the other two phases.  In the slow-roll phase, $w(\phi)$ approaches to $-1$, and remains constant  till the end of the slow-roll inflation.
It is very interesting to note that the evolution of $a(t)$ (the left panel of Fig. \ref{fig:mod1}) during the bouncing phase is universal, and shows consistent behavior with the analytical solution (\ref{eq:a}).

The  range of the initial conditions is $\phi_B \in(\phi_{min}, \infty)$, in which  the KED condition at the bounce is assured, as in this range $\dot{\phi_B}^2/2\gg V(\phi_B)$ is always true, and 
 it always leads to a slow-roll inflationary phase. Next, we turn to consider  the total number of $e$-folds during the slow-roll inflation for various values of  
$\phi_B$. To be consistent with the Planck 2018 results \cite{Planck2018}, at least 60 $e$-folds are required for a successful inflationary model. However, in the case  $\alpha = 1 m_{Pl}^2$ and  $c = 8.343 \times 10^{-7} m_{Pl}$ 
the  $e$-folds are less than 60, which are shown in Table \ref{tab:mod1} for different values of $\phi_B$.

We also analyzed the case with $\alpha = 0.5 m_{Pl}^2$ and $c = 1.611 \times 10^{-6} m_{Pl}$, and noticed that the conclusion is the same. In fact, as we mentioned previously, we found that this is true for all the sets of  $(\alpha, c)$ that satisfy the Planck 2018 data. So,  in order not to repeat the calculations, we do not present the detailed analyses for this case, as well as the other ones.

\begin{table}[tbp]
\caption{This table represents model 1 [Eq.(\ref{eq:pot1})] with $\dot{\phi}_B > 0$. We demonstrate  various parameters of  inflation for different values of $\phi_B$ in the case of $\alpha = 1 m_{Pl}^2$ and  $c = 8.343 \times 10^{-7} m_{Pl}$. For each value of $\phi_B$, we get less than 60 $e$-folds. Therefore, these initial values of $\phi_B$ are not consistent with observations.}
\begin{center}
\resizebox{\textwidth}{!}{
\begin{scriptsize}
\begin{tabular}{cccccccc}
\hline
$\phi_B/m_{Pl}$~~~  & Inflation~~~ & $t/t_{Pl}$~~~ & $\epsilon$~~ & $w$ ~~& $N_{inf}$ &~~~${w}^B$\\
\hline
0.01 ~~~& begin~~~& $1.17480 \times 10^5$~~~& 1.0~~ & $-1/3$ ~~& ~~~& ~~~&\\
& slow-roll~~~& $2.84048 \times 10^5$ ~~~& 0.073~~ & $-0.950$ ~~& 31.38 ~~~& $>0$\\
& end~~~& $1.0407 \times 10^7$ ~~~& 0.174~~ & $-1/3$ ~~& ~~~& ~~~& \\\\
1 ~~~& begin~~~& $1.39037 \times 10^5$~~~& 0.999~~ & $-1/3$ ~~& ~~~& ~~~&\\
& slow-roll~~~& $3.18522 \times 10^5$ ~~~& 0.074~~ & $-0.950$ ~~& 27.70 ~~~& $>0$\\
& end~~~& $1.0371 \times 10^7$ ~~~& 0.149~~ & $-1/3$ ~~& ~~~& ~~~& \\\\
10 ~~~& begin~~~& $1.58197 \times 10^5$~~~& 0.999~~ & $-1/3$ ~~& ~~~& ~~~&\\
& slow-roll~~~& $3.50170 \times 10^5$ ~~~& 0.074~~ & $-0.950$ ~~& 25.37 ~~~& $>0$\\
& end~~~& $1.0687 \times 10^7$ ~~~& 0.218~~ & $-1/3$ ~~& ~~~& ~~~& \\
\hline
\end{tabular}
\end{scriptsize}}
\label{tab:mod1}
\end{center}
\end{table}

\subsection{Model 2}
\label{subsec:mod2}

In this subsection, we study some characteristics of model 2 [Eq.(\ref{eq:pot2})], for which the potential  is displayed in the upper right panel of Fig. \ref{fig:pot}. In the large field limit ($\phi \rightarrow \infty$), the potential monotonically declines to a finite value $V(\phi) \rightarrow \alpha c^2/4$,  whereas at $\phi \rightarrow -\infty$, it diverges. In LQC, $\rho_c$ constraints the value of $\phi_B$ as $(\phi_{min}, \infty)$, and $\phi_{min}$ is given by
\begin{eqnarray}
\phi_{min} &\simeq & -\sqrt{\frac{3\alpha}{2}}~ \text{Log} \left( \sqrt{\frac{4\rho_c}{\alpha c^2}}-1 \right).
\label{eq:mod2phimin}
\end{eqnarray}

\begin{figure}[tbp]
\begin{center}
\begin{tabular}{ccc}
{\includegraphics[width=1.9in,height=1.65in,angle=0]{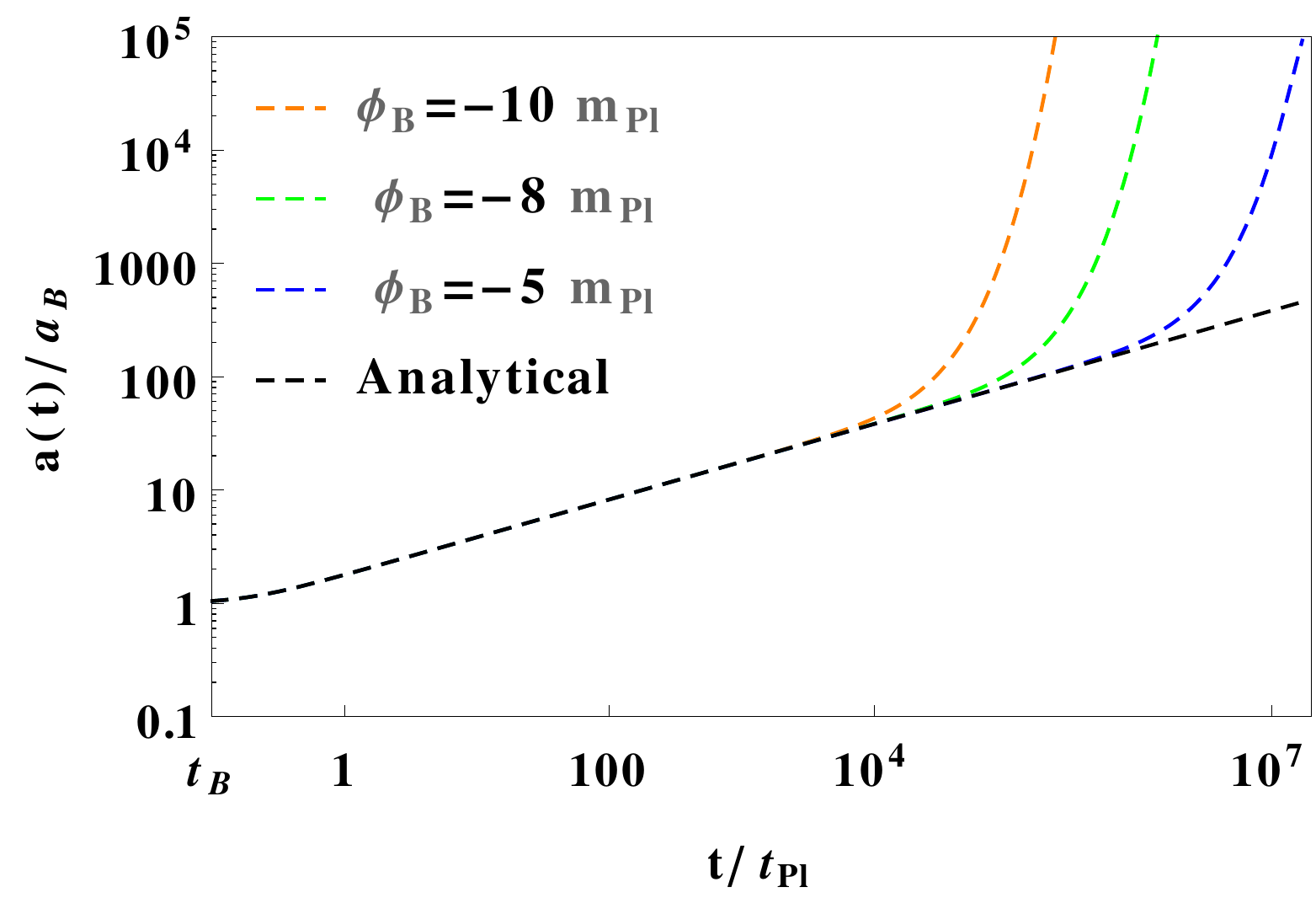}} &
{\includegraphics[width=1.9in,height=1.6in,angle=0]{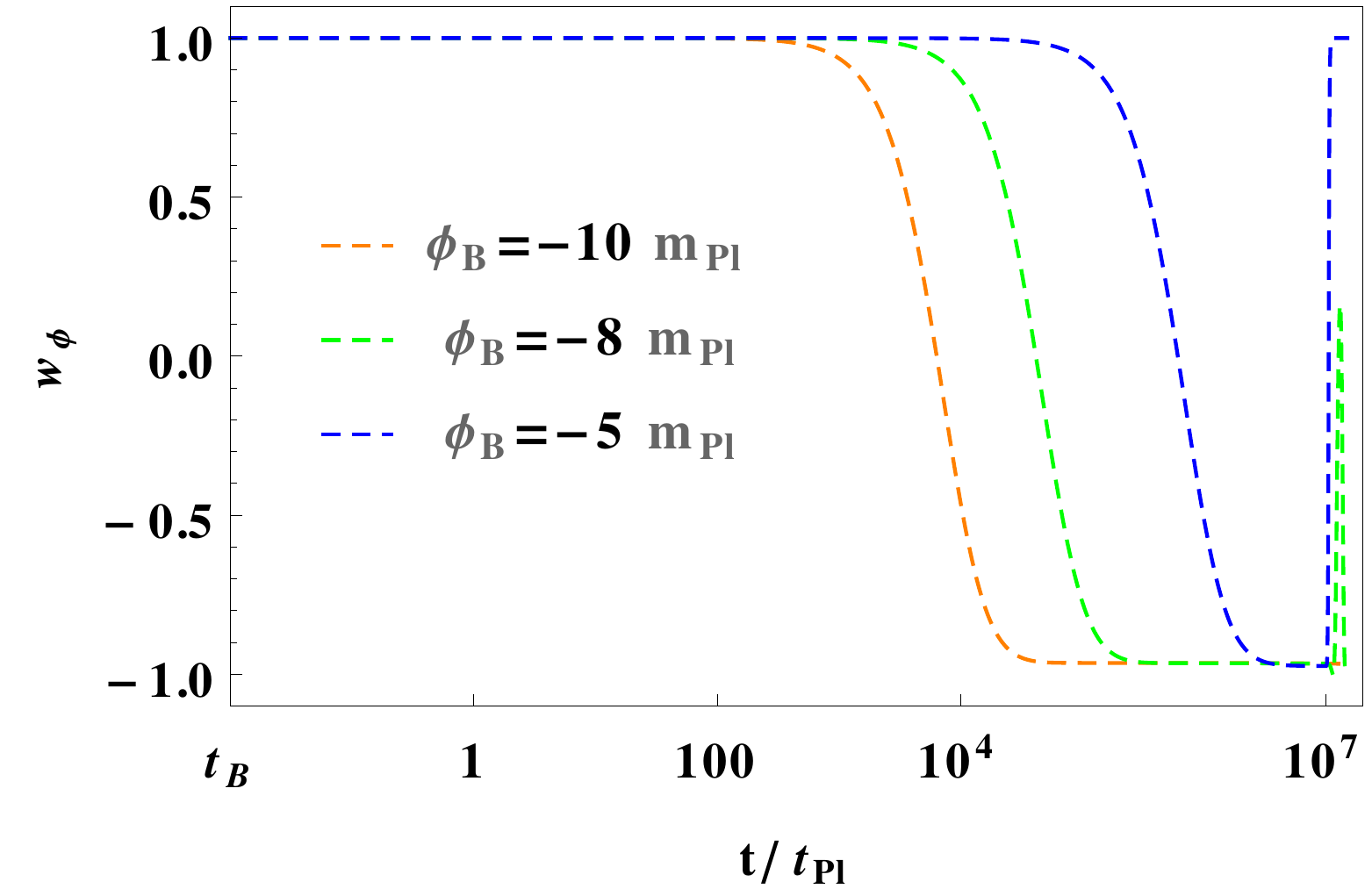}} &
{\includegraphics[width=1.9in,height=1.6in,angle=0]{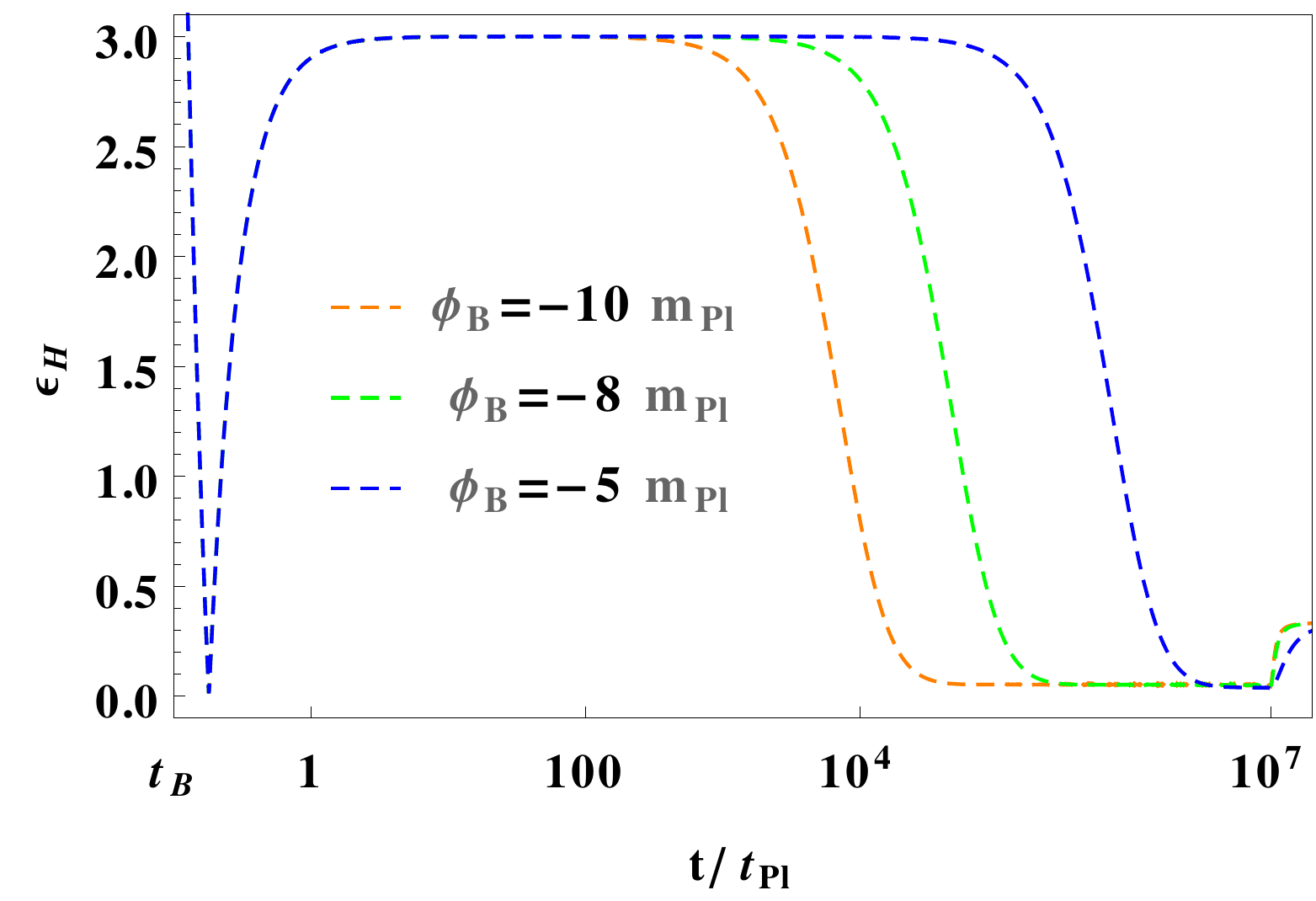}}
\\
{\includegraphics[width=1.9in,height=1.65in,angle=0]{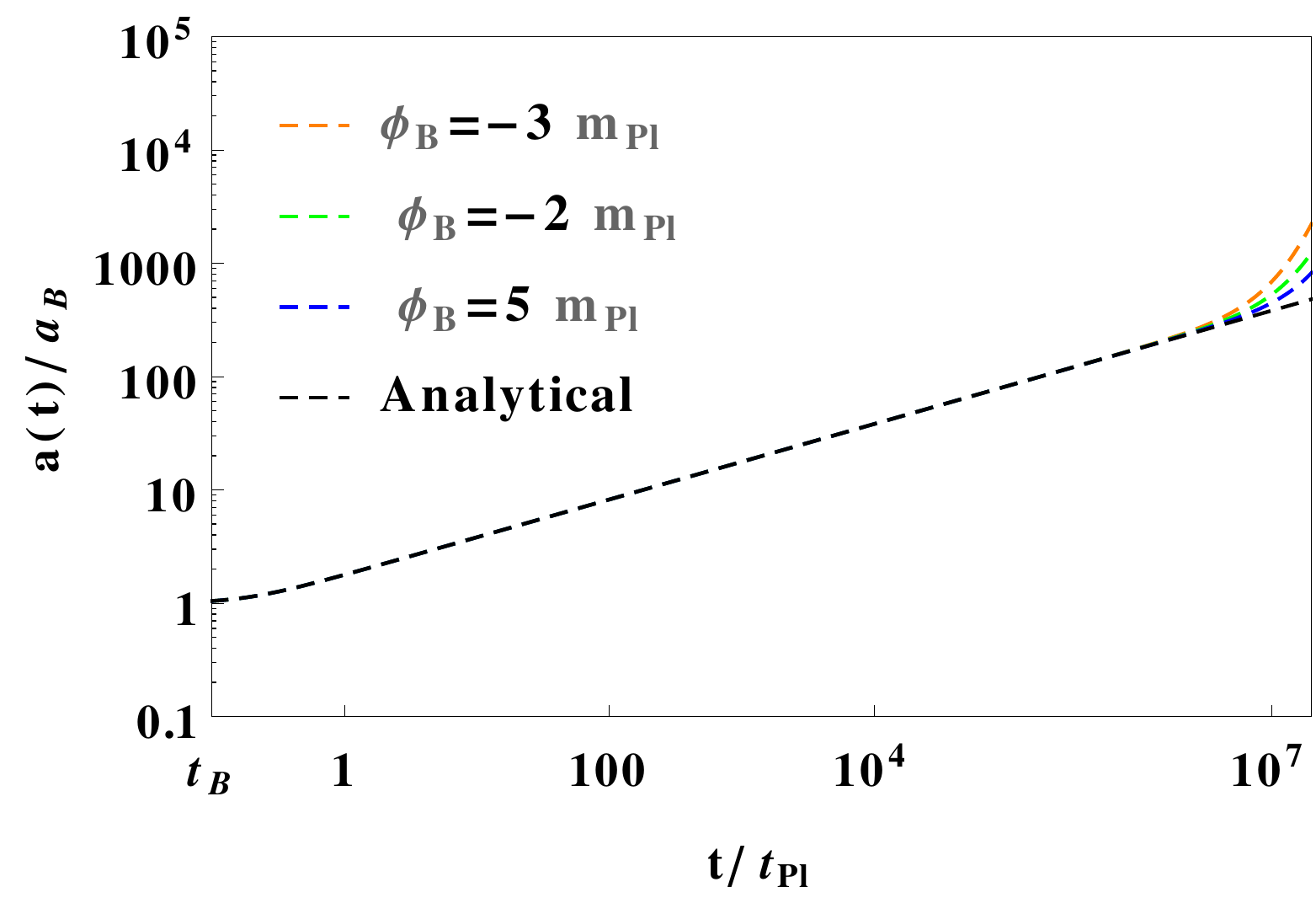}} &
{\includegraphics[width=1.9in,height=1.6in,angle=0]{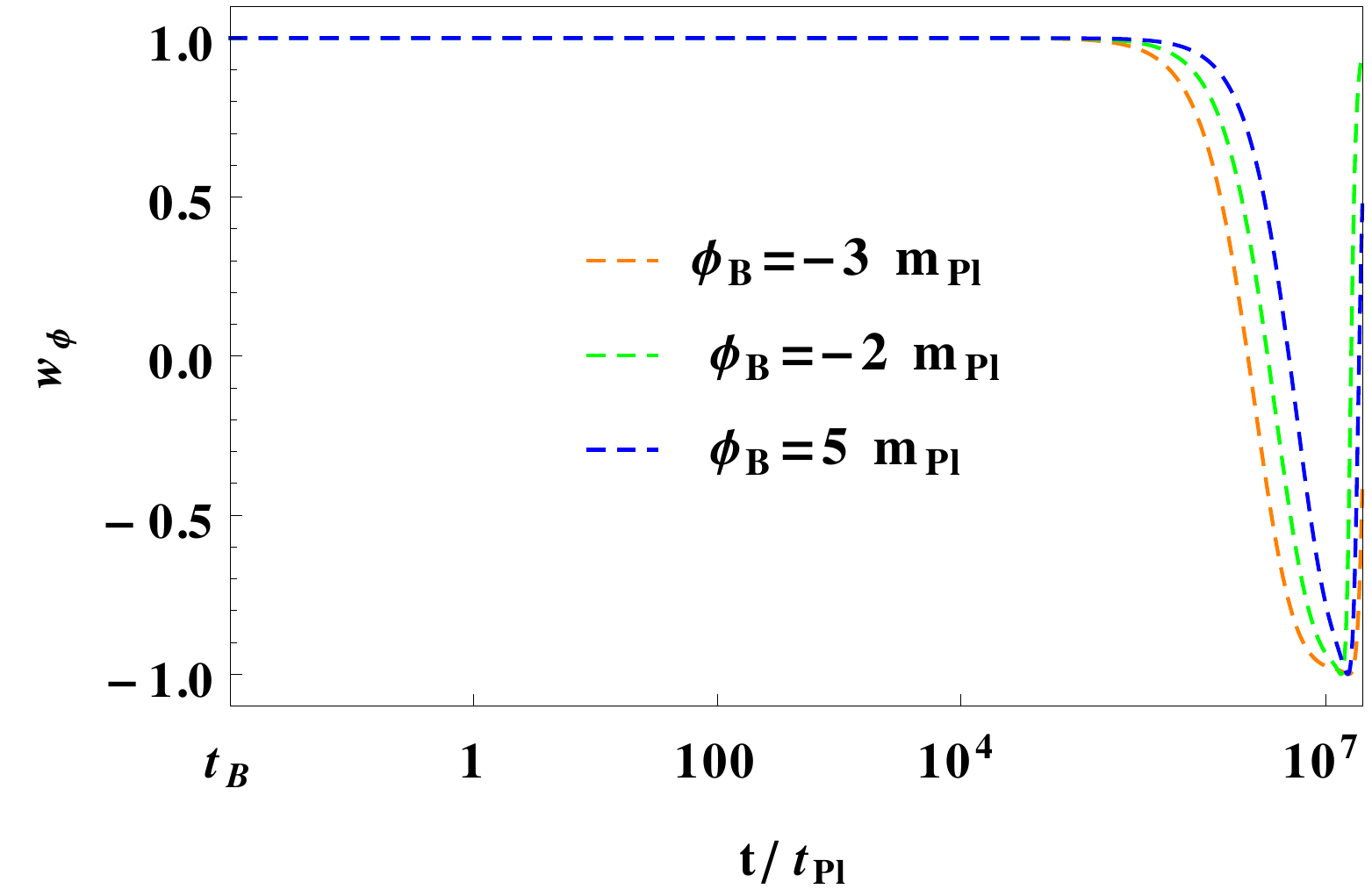}} &
{\includegraphics[width=1.9in,height=1.6in,angle=0]{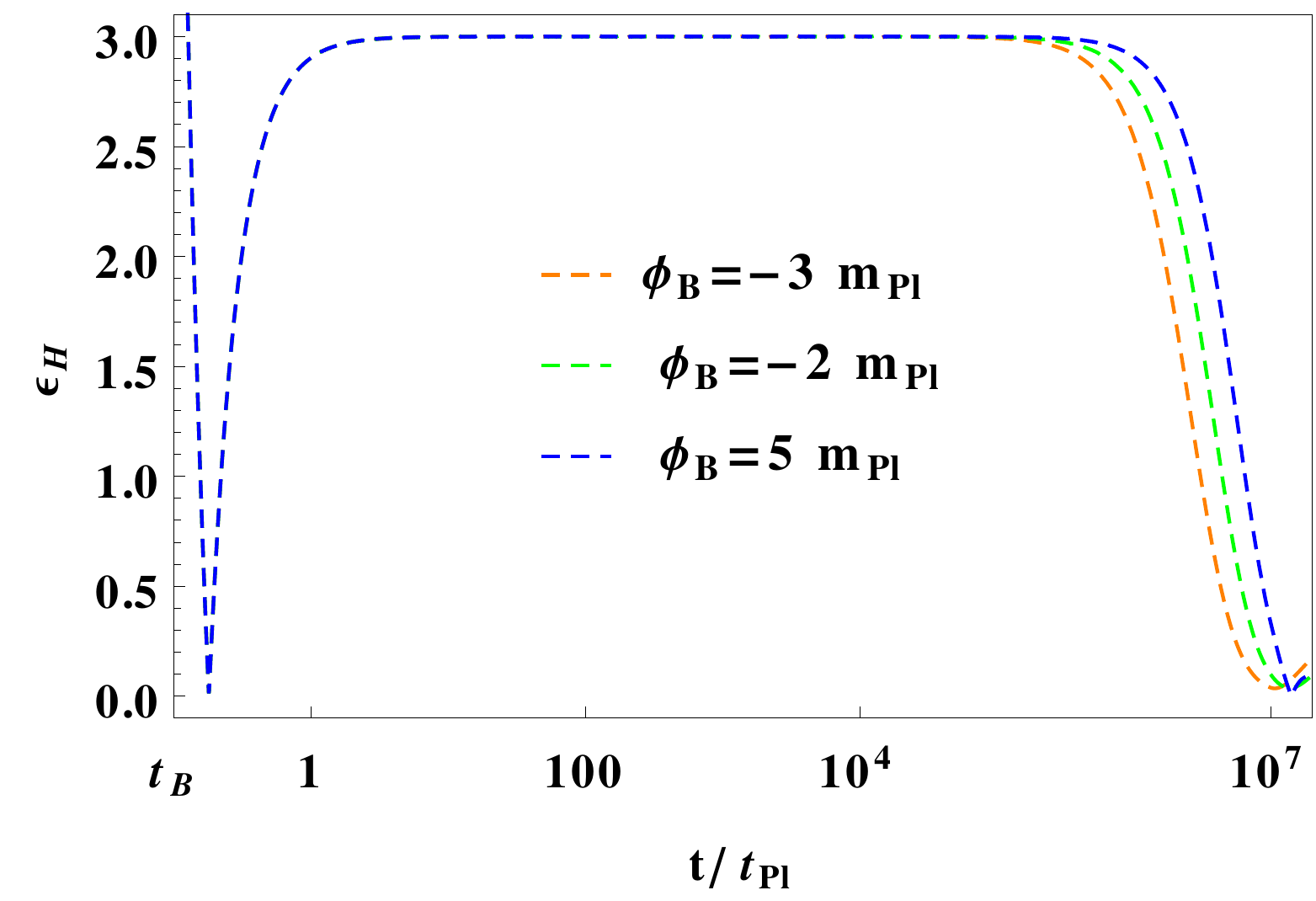}}
\\
{\includegraphics[width=1.9in,height=1.6in,angle=0]{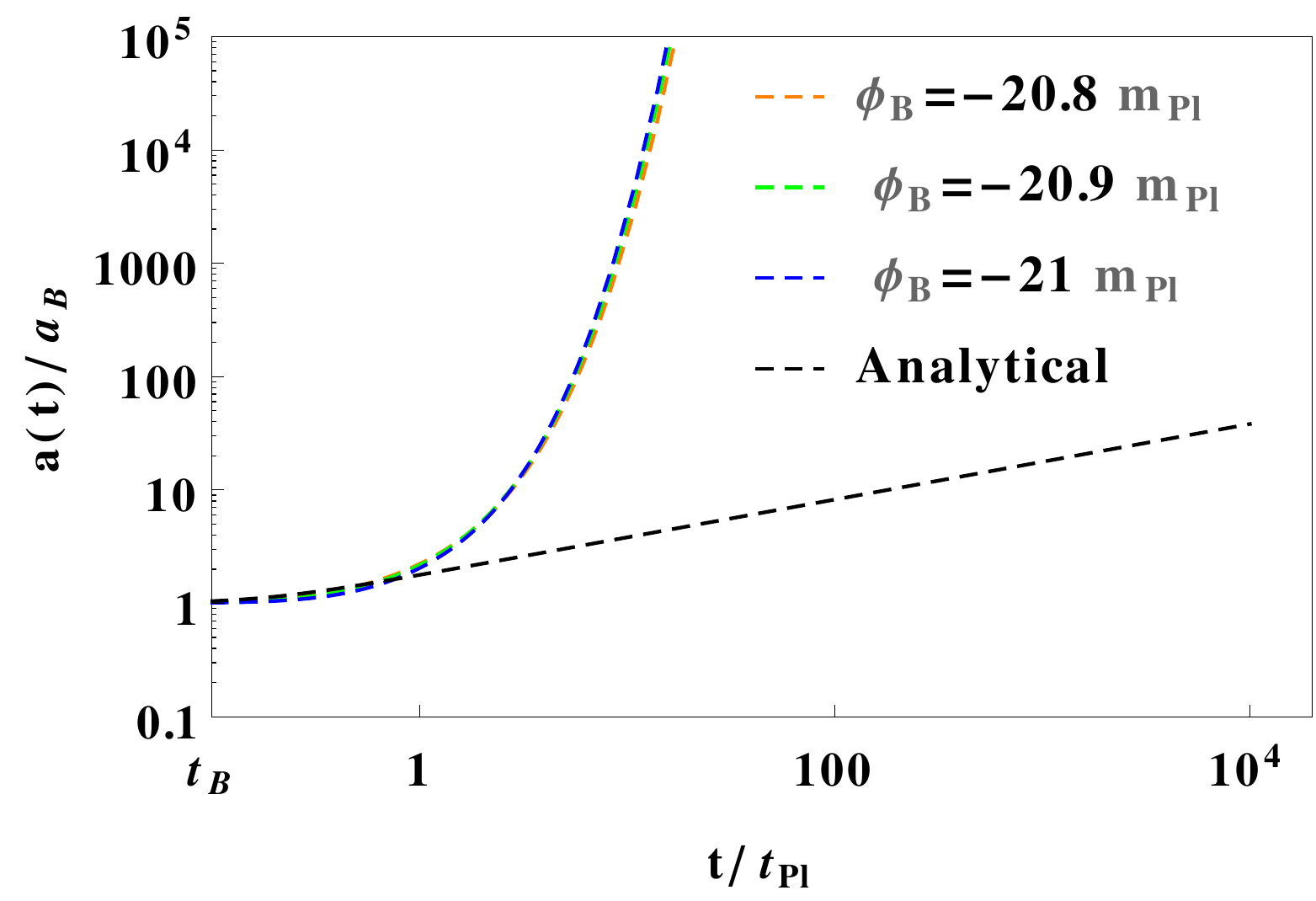}} & 
{\includegraphics[width=1.9in,height=1.6in,angle=0]{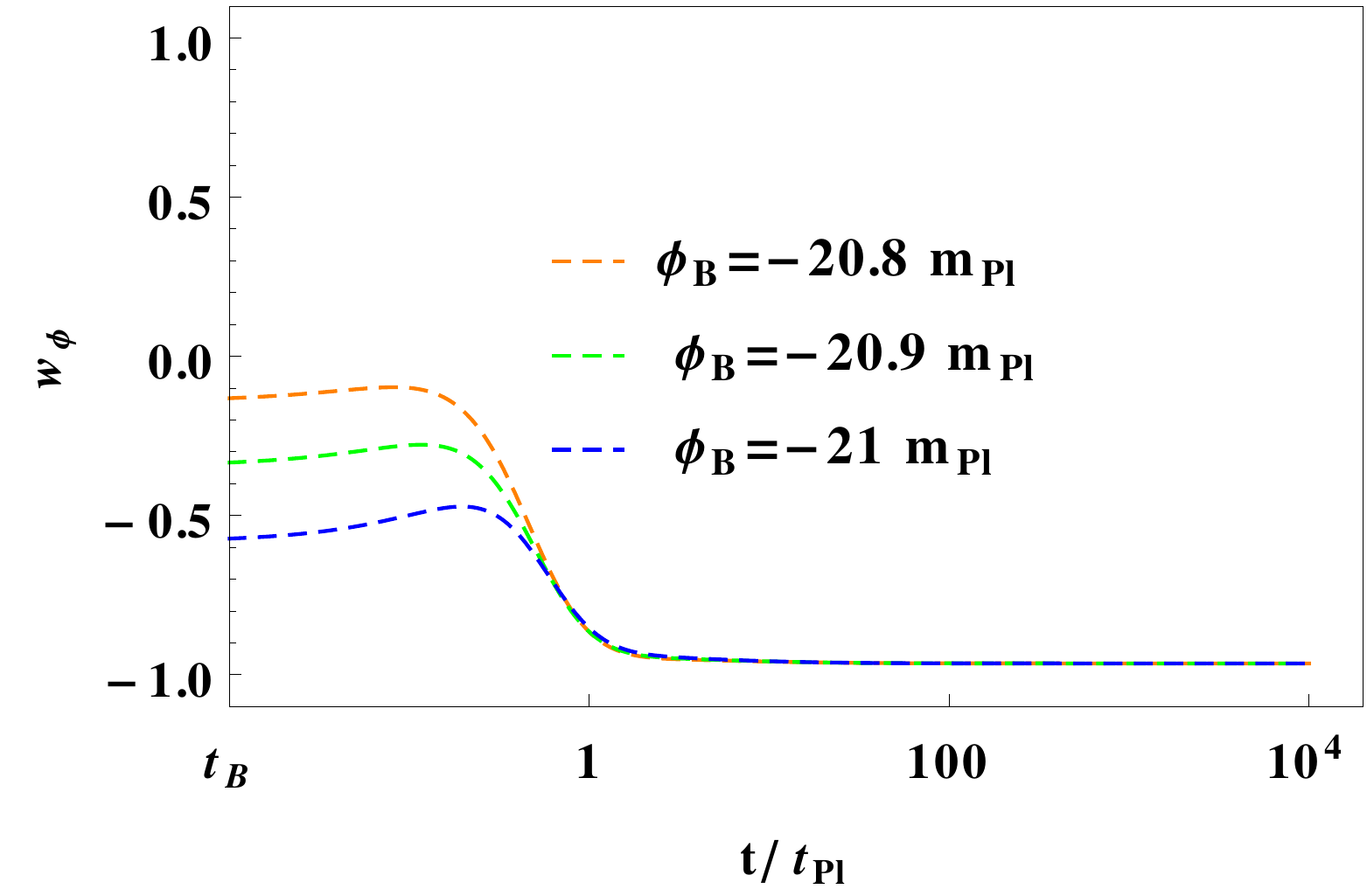}} & 
{\includegraphics[width=1.9in,height=1.6in,angle=0]{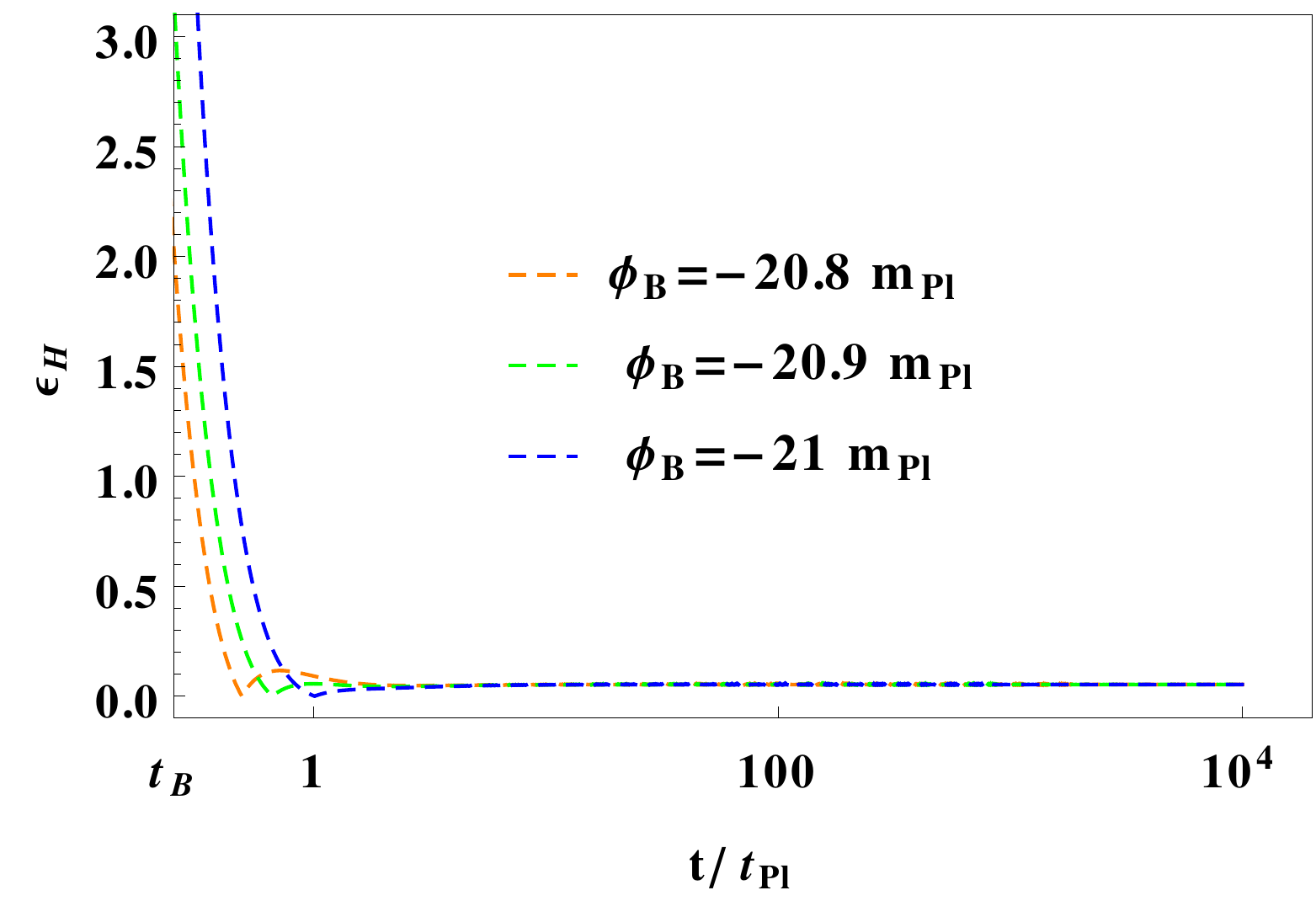}} 
\end{tabular}
\end{center}
\caption{This figure demonstrates the numerical evolution of $a(t)$, $w(\phi)$ and $\epsilon_H$ for model 2 [Eq.(\ref{eq:pot2})] with $\dot{\phi_B}>0$. Top (KED) and bottom (PED) panels provide the slow-roll inflationary phase, whereas a subset of the KED initial conditions (middle panels) do not lead to the slow-roll inflation. When plotting out the figure,  we had set $\alpha = 1 m_{Pl}^2$,  $c = 4.074 \times 10^{-8} m_{Pl}$ and $m_{Pl}=1$.  }
\label{fig:mod2}
\end{figure}

\begin{table}[tbp]
\caption{This table corresponds to model 2 [Eq.(\ref{eq:pot2})] with $\dot{\phi}_B > 0$. We show the number of $e$-folds $N_{inf}$ and other parameters of inflation for different choices of $\phi_B$ with the set of $\alpha = 1 m_{Pl}^2$ and  $c = 4.074 \times 10^{-8} m_{Pl}$.}
\begin{center}
\resizebox{\textwidth}{!}{
\begin{scriptsize}
\begin{tabular}{cccccccccc}
\hline
$\phi_B/m_{Pl}$~~~  & Inflation~~~ & $t/t_{Pl}$~~~ & $\epsilon$~~ & $w$ ~~& $N_{inf}$ &~~~${w}^B$\\
\hline
$-20.9$ ~~~& begin~~~& 0.01 ~~~& 3.17~~ & $-1/3$ ~~& ~~~& ~~~&\\
& slow-roll~~~& 2.06 ~~~& 0.043~~ & $-0.978$ ~~& 254.98 ~~~&$<0$\\
& end~~~& $1.735 \times 10^7$ ~~~& 0.329~~ & $-1/3$ ~~& ~~~& ~~~& \\\\
$-10$ ~~~& begin~~~& $8.764 \times 10^3$ ~~~& 0.999~~ & $-1/3$ ~~& ~~~& ~~~&\\
& slow-roll~~~& 4.4992$\times 10^4$ ~~~& 0.057~~ & $-0.961$ ~~& 68.25 ~~~&$>0$\\
& end~~~& $2.432 \times 10^7$ ~~~& 0.332~~ & $-1/3$ ~~& ~~~& ~~~& \\\\
$-9.7$ ~~~& begin~~~& $1.1622 \times 10^4$ ~~~& 0.999~~ & $-1/3$ ~~& ~~~& ~~~&\\
& slow-roll~~~& 7.6452$\times 10^4$ ~~~& 0.054~~ & $-0.964$ ~~& 60.45 ~~~& $>0$\\
& end~~~& $1.541 \times 10^7$ ~~~& 0.326~~ & $-1/3$ ~~& ~~~& ~~~& \\\\
$-9$ ~~~& begin~~~& $2.2426 \times 10^4$ ~~~& 0.999~~ & $-1/3$ ~~& ~~~& ~~~&\\
& slow-roll~~~& 1.10808$\times 10^5$ ~~~& 0.058~~ & $-0.961$ ~~&50.68 ~~~&$>0$\\
& end~~~& $2.066 \times 10^7$ ~~~& 0.331~~ & $-1/3$ ~~& ~~~& ~~~& \\
\hline
\end{tabular}
\end{scriptsize}}
\label{tab:mod2}
\end{center}
\end{table}

To find the values of $\alpha$ and $c$ that are  consistent  with the Planck 2018 data \cite{Planck2018}, following what is prescribed in Appendix A,  we find various sets of $\alpha$ and $c$, similar to Model 1. In the current model, it is sufficient to consider only the following two representative cases, 
\begin{eqnarray}
\alpha &=& 1 m_{Pl}^2, \qquad~~~~ c = 4.074 \times 10^{-8} m_{Pl}\nonumber \\
\alpha &=& 5 m_{Pl}^2, \qquad~~~~ c = 2.449 \times 10^{-7} m_{Pl}.
\label{eq:mod2alphac}
\end{eqnarray}
The value of $\phi_{min}$ can be obtained for any choice of $\alpha$ and $c$ from Eq. (\ref{eq:mod2phimin}). For example, for $\alpha = 1 m_{Pl}^2$ and  $c = 4.074 \times 10^{-8} m_{Pl}$, we find  $\phi_{min}=-21.14 m_{Pl}$. We investigate the entire range of inflaton field in order to identify the initial values that provide the slow-roll inflation.
\begin{eqnarray}
\frac{\phi_B}{m_{Pl}}  =\begin{cases}
\in  (\phi_{min}, -20.73), &  \text{PED (slow-roll)},\cr  
= -20.72 , &  \text{KE=PE (slow-roll)}, \cr 
\in  (-20.71,-3.5), &   \text{KED (slow-roll)}, \cr  
\in (-3.4, \infty), & \text{KED (no slow-roll)},\cr
\end{cases}
\label{eq:mod2phiB}
\end{eqnarray}
where
$\phi_{min}$ is given by Eq. (\ref{eq:mod2phimin}). The results of background evolution for KED and PED initial conditions are exhibited in Fig. \ref{fig:mod2} with various choices of   $\phi_B$. In the  KED case, the evolution of $a(t)$ shows the universal feature during the bouncing phase, that is,  it neither depends on potential nor on the initial values of $\phi_B$, and is well described by the analytical solution (\ref{eq:a}). This is because  during the whole
phase,   the potential remains almost constant, and does not essentially affect the evolution of the background. From the evolution of $w(\phi)$, one can see that in the KED case the background evolution is split up into three different phases: bouncing, transition and slow-roll. The period of the transition phase is very short in comparison with the other two phases. During the bouncing regime, $w(\phi) \simeq +1$, in the transition regime, it decreases drastically from $+1$ $(t/t_{Pl} \simeq 10^4)$ to $-1$ $(t/t_{Pl} \simeq 10^6)$, and in the slow-roll regime $w(\phi) \simeq -1$ until the end of slow-roll inflation. In the case of KED initial conditions, we also find a subset where the slow-roll inflation is not possible,  which is clearly displayed in the middle panels of Fig. \ref{fig:mod2}. In the PED case, the universality of $a(t)$ disappears, and the bouncing and transition phases do not exist any more, however the slow-roll inflation can still be obtained as shown in lower panels of   Fig. \ref{fig:mod2}.

Table \ref{tab:mod2} shows various parameters of inflation. In particular,   $N_{inf}$ decreases as  $\phi_B$ grows. From this table, one can find the range of $\phi_B$ that provides 60 or more $e$-folds to be compatible with observations, which is 
\begin{eqnarray}
\frac{\phi_B}{m_{Pl}}  &\in&  (\phi_{min}, -9.7), \; \;\; N_{inf} \gtrsim 60, 
\label{eq:mod2phiB60}
\end{eqnarray}
 where $\phi_{min}$ is given by Eq. (\ref{eq:mod2phimin}). 

We also examined the other set of Eq. (\ref{eq:mod2alphac}), namely $\alpha = 5 m_{Pl}^2$ and  $c = 2.449 \times 10^{-7} m_{Pl}$, and observed that the subset of the KED case, which  does not provide an inflationary phase found in the case of $\alpha=1 m_{Pl}^2$, disappears. In fact, we found that this is true for all the cases with a large enough value of $\alpha$.  Therefore, we conclude that the entire range of KE and PE at the bounce provides inflationary phase. Though, a portion of this entire range  provides less than 60 $e$-folds. Similar to Eq. (\ref{eq:mod2phiB60}), in this case, we shall also get restricted range of the inflaton field that is consistent with current observations. Moreover, the results are highly depend on the values of $\alpha$ and $c$.

\begin{figure}[tbp]
\begin{center}
\begin{tabular}{ccc}
{\includegraphics[width=1.9in,height=1.65in,angle=0]{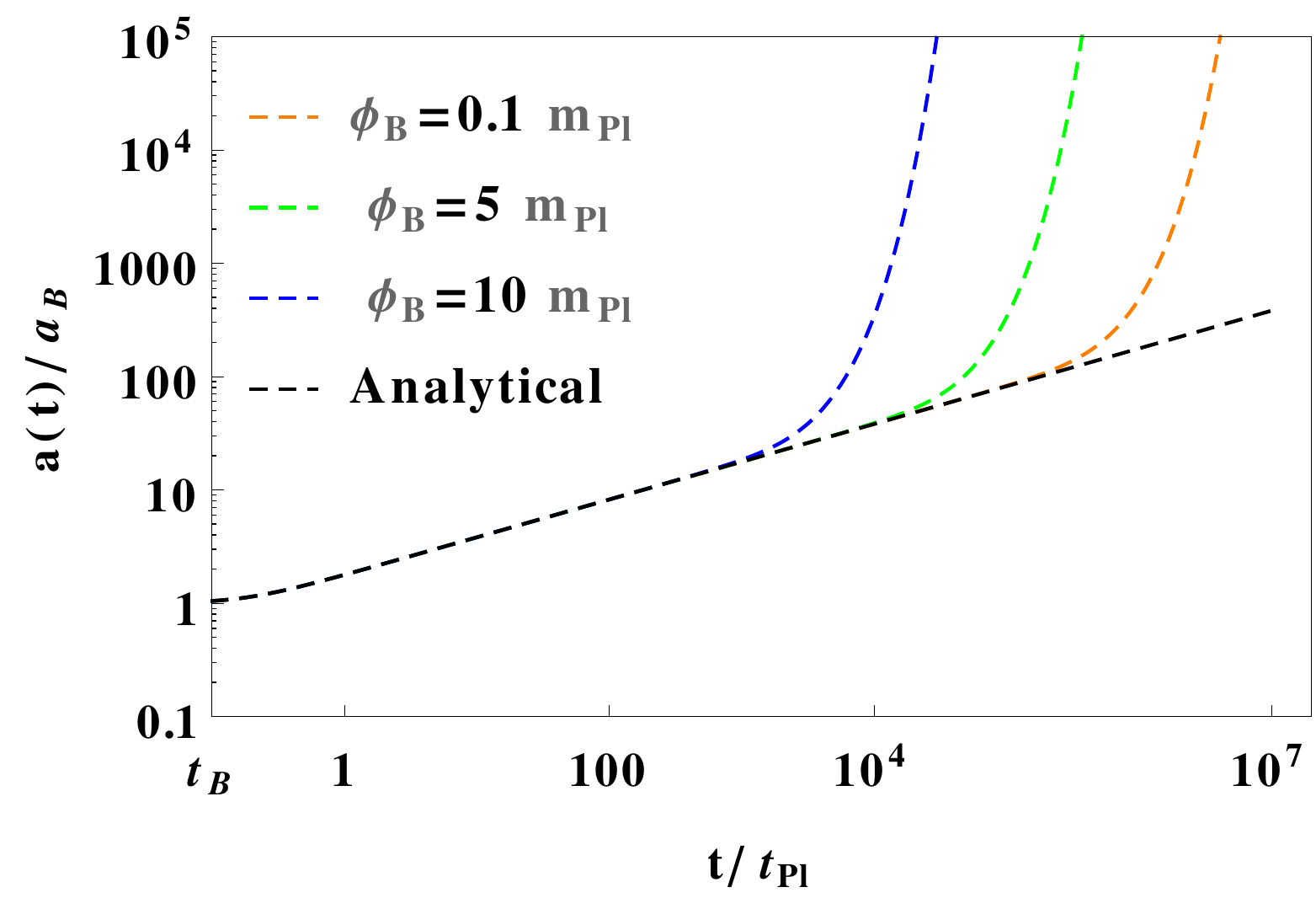}} &
{\includegraphics[width=1.9in,height=1.6in,angle=0]{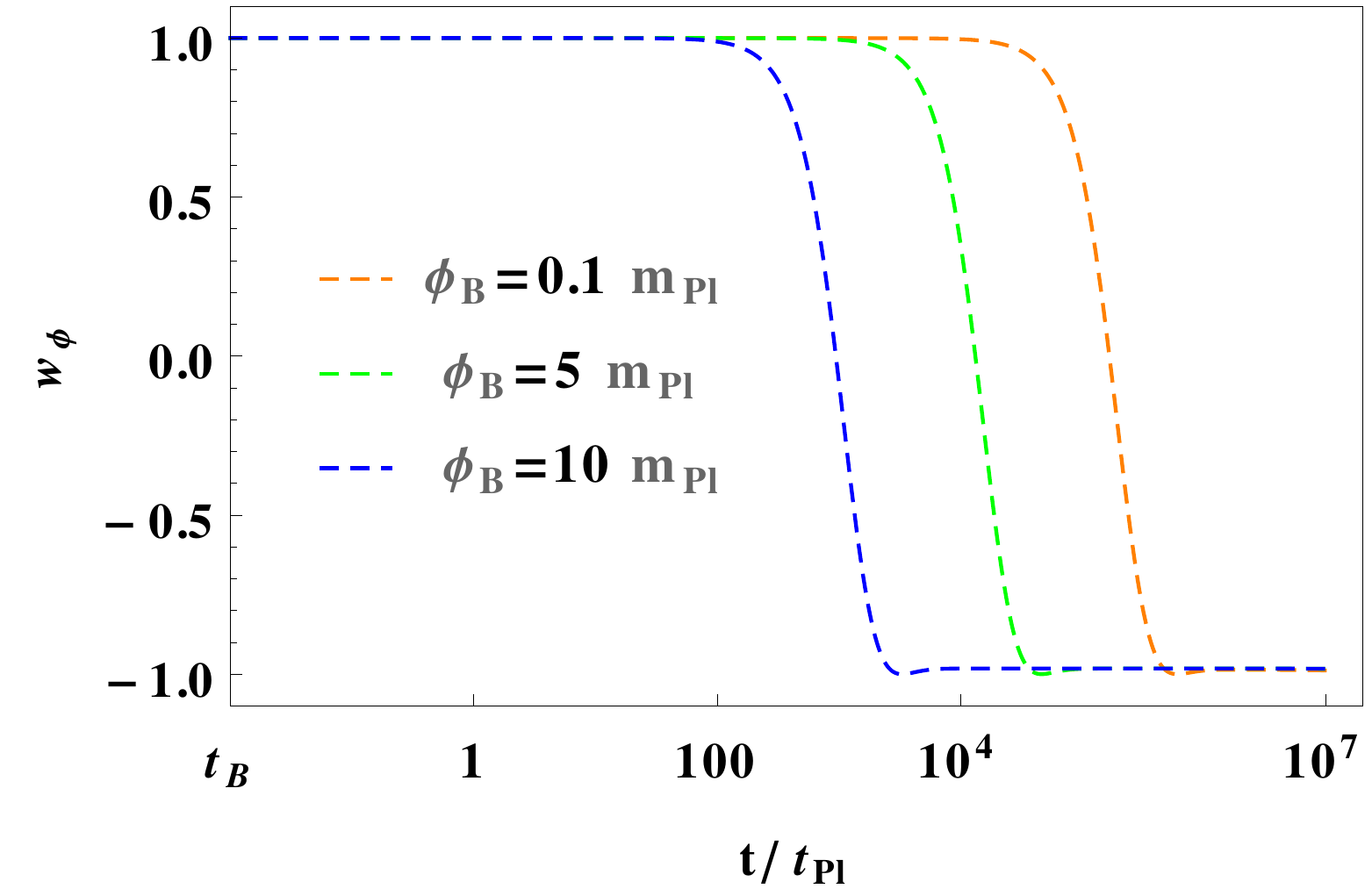}} &
{\includegraphics[width=1.9in,height=1.6in,angle=0]{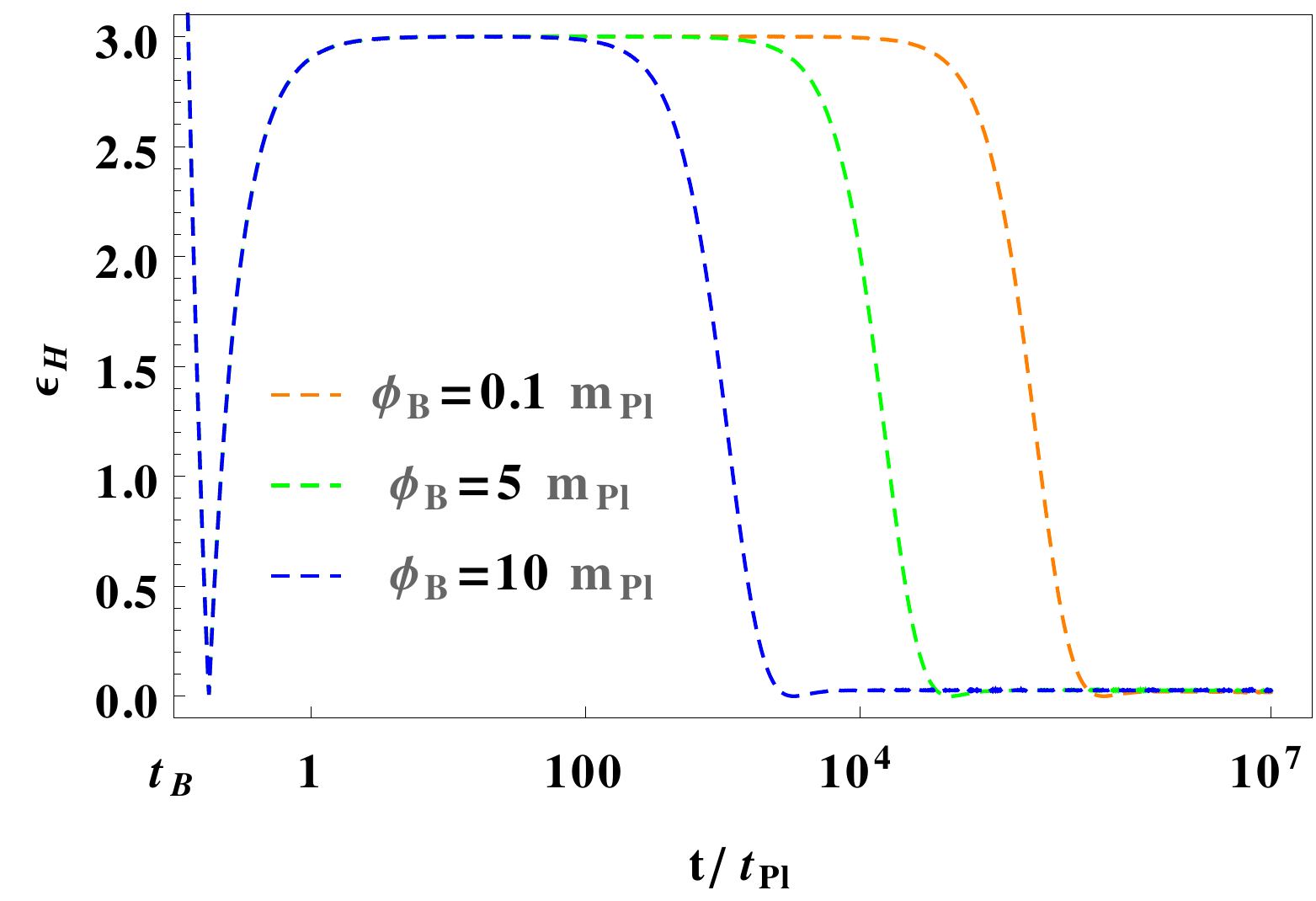}}
\\
{\includegraphics[width=1.9in,height=1.6in,angle=0]{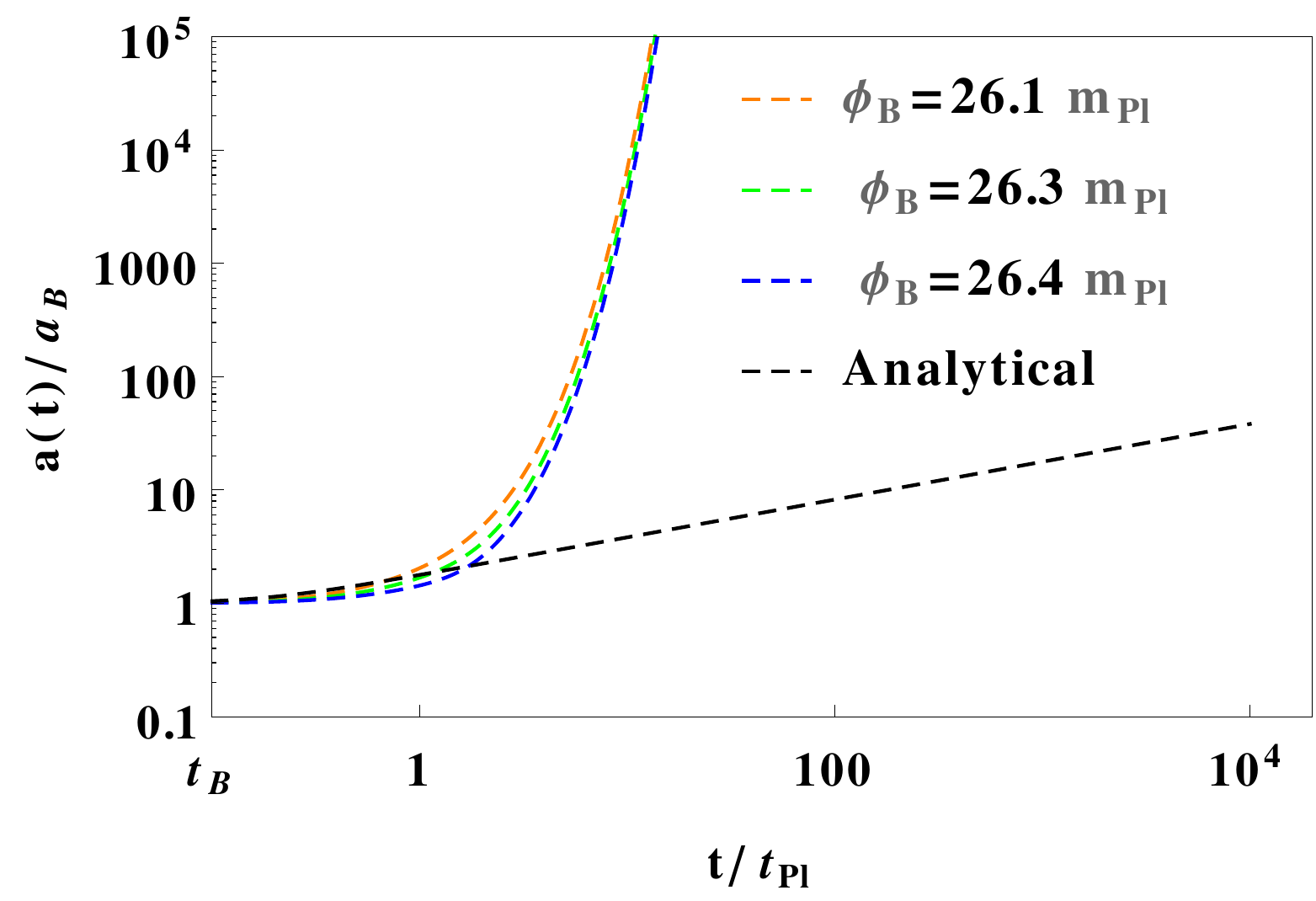}} & 
{\includegraphics[width=1.9in,height=1.6in,angle=0]{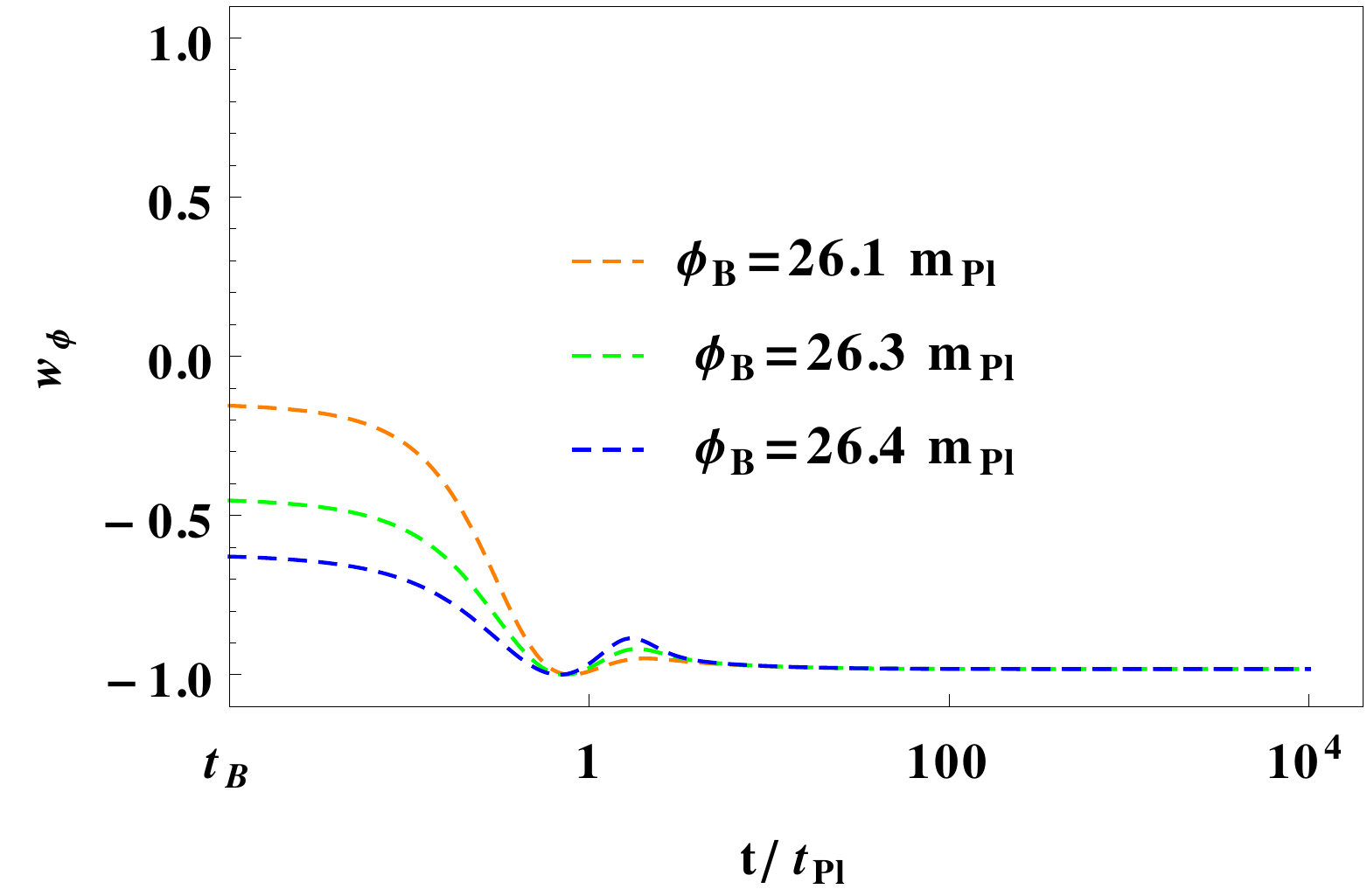}} & 
{\includegraphics[width=1.9in,height=1.6in,angle=0]{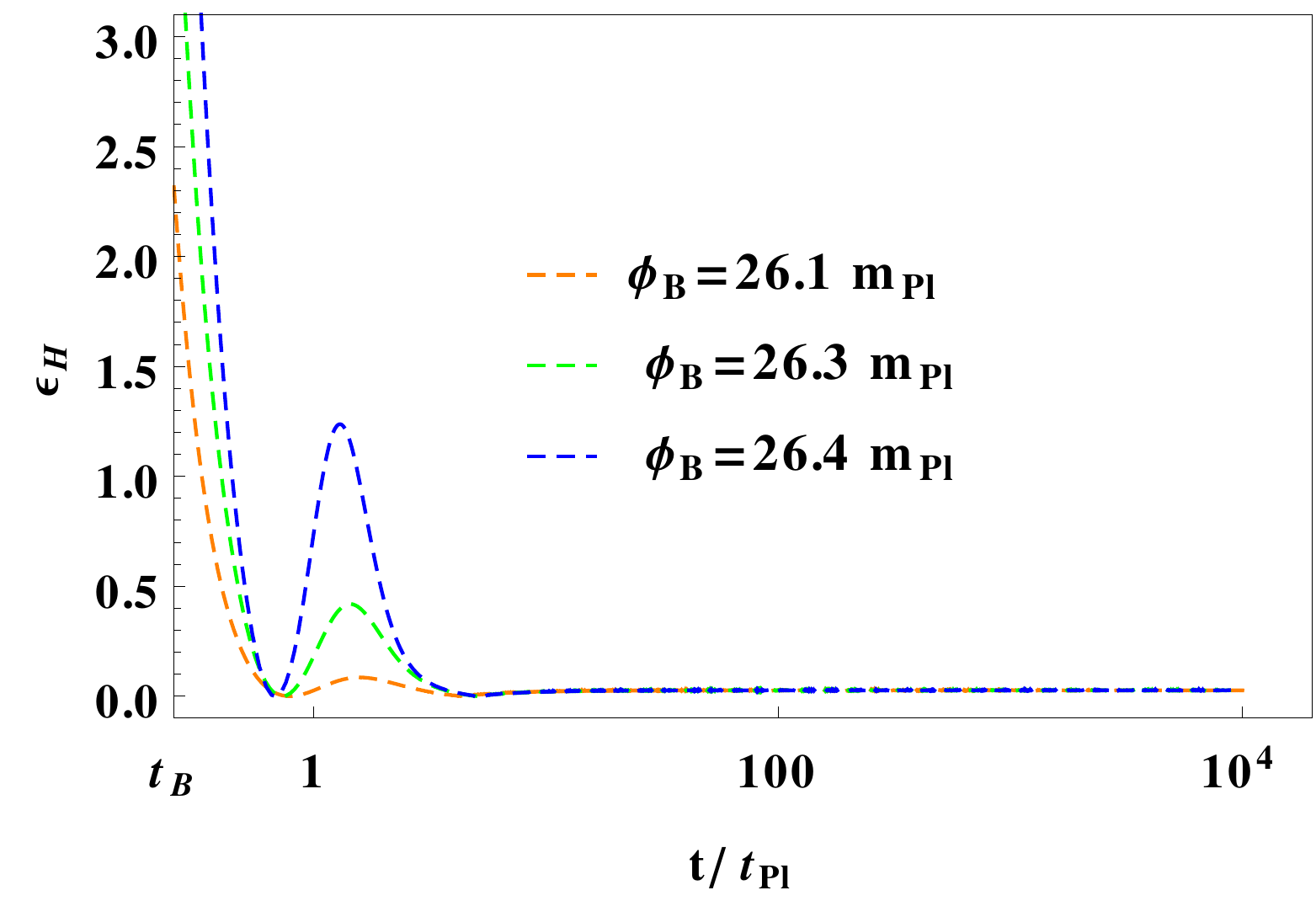}} 
\end{tabular}
\end{center}
\caption{This figure is for model 3 [Eq.(\ref{eq:pot3})] with $\dot{\phi_B}>0$. The potential (\ref{eq:pot3}) is symmetric with respect to $\phi=0$. Therefore, one can get similar results for $\dot{\phi_B}<0$. In the entire range of the initial conditions of $\phi_B$ (top: KED and bottom: PED), the slow-roll inflation is always obtained. When plotting out the figure,  we had set  
 $\alpha = 0.5 m_{Pl}^2$ and  $c = 3.915 \times 10^{-7} m_{Pl}$ and $m_{Pl}=1$.  }
\label{fig:mod3}
\end{figure}

\begin{table}[tbp]
\caption{This table designates the model 3 [Eq.(\ref{eq:pot3})] with $\dot{\phi}_B > 0$, and $\alpha = 0.5 m_{Pl}^2$ and  $c = 3.915 \times 10^{-7} m_{Pl}$.}
\begin{center}
\resizebox{\textwidth}{!}{
\begin{scriptsize}
\begin{tabular}{cccccccccc}
\hline
$\phi_B/m_{Pl}$~~~  & Inflation~~~ & $t/t_{Pl}$~~~ & $\epsilon$~~ & $w$ ~~& $N_{inf}$ &~~~${w}^B$\\
\hline
$26.3$ ~~~& begin~~~& 0.01 ~~~& 2.55~~ & $-1/3$ ~~& ~~~& ~~~&\\
& slow-roll~~~& 4.5 ~~~& 0.007~~ & $-0.960$ ~~& 485.59 ~~~&$<0$\\
& end~~~& $2.38 \times 10^7$ ~~~& 0.333~~ & $-1/3$ ~~& ~~~& ~~~& \\\\
6 ~~~& begin~~~& 9.85948 $\times 10^3$~~~& 1.0~~ & $-1/3$ ~~& ~~~& ~~~&\\
& slow-roll~~~& 2.73168$ \times10^4$ ~~~& 5.02677$ \times10^{-6}$~~ & $-1$ ~~& 99.38 ~~~&$>0$\\
& end~~~& $1.371 \times 10^7$ ~~~& 0.329~~ & $-1/3$ ~~& ~~~& ~~~& \\\\
3.75 ~~~& begin~~~& 3.23086 $\times 10^4$~~~& 0.999~~ & $-1/3$ ~~& ~~~& ~~~&\\
& slow-roll~~~& 8.95696$ \times10^4$ ~~~& 2.42$ \times10^{-5}$~~ & $-1$ ~~& 60.21 ~~~&$>0$\\
& end~~~& $1.2465 \times 10^7$ ~~~& 0.322~~ & $-1/3$ ~~& ~~~& ~~~& \\\\
3 ~~~& begin~~~& 4.79092 $\times 10^4$~~~& 0.999~~ & $-1/3$ ~~& ~~~& ~~~&\\
& slow-roll~~~& 1.32921$ \times10^5$ ~~~& 6.85$ \times10^{-5}$~~ & $-1$ ~~& 49.48 ~~~&$>0$\\
& end~~~& $1.337 \times 10^7$ ~~~& 0.325~~ & $-1/3$ ~~& ~~~& ~~~& \\\\
$-8.22$ ~~~& begin~~~& 2.77706 $\times 10^4$~~~& 1.0~~ & $-1/3$ ~~& ~~~& ~~~&\\
& slow-roll~~~& 1.40387$ \times10^5$ ~~~& 3.0$ \times10^{-2}$~~ & $-0.98$ ~~& 60.15 ~~~&$>0$\\
& end~~~& $1.0837 \times 10^7$ ~~~& 0.266~~ & $-1/3$ ~~& ~~~& ~~~& \\\\
$-9$ ~~~& begin~~~& 1.69003 $\times 10^4$~~~& 0.99~~ & $-1/3$ ~~& ~~~& ~~~&\\
& slow-roll~~~& 6.14094$ \times10^4$ ~~~& 4.5$ \times10^{-2}$~~ & $-0.97$ ~~& 76.56 ~~~&$>0$\\
& end~~~& $1.1434 \times 10^7$ ~~~& 0.308~~ & $-1/3$ ~~& ~~~& ~~~& \\
\hline
\end{tabular}
\end{scriptsize}}
\label{tab:mod3}
\end{center}
\end{table}

\subsection{Model 3}
\label{subsec:mod3}

In this subsection, let us consider potential (\ref{eq:pot3}) (model 3). The evolution of this potential is exhibited in the lower left panel of Fig. \ref{fig:pot}. The potential is symmetric with respect to  $\phi=0$, bounded below by unity ($V(\phi) \geq 1$), and shows oscillations as the field approaches to the origin ($\phi \rightarrow 0$). In the large field limit ($\phi \rightarrow \pm \infty$), the potential is unbounded, and the maximum energy density $\rho_c$ restricts the range of $\phi_B$ to $(\phi_{min}, \phi_{max})$, where 
\begin{eqnarray}
\phi_{max, \; min} &\simeq & \pm \sqrt{6 \alpha}~ \text{arccosh} \left( \sqrt{\frac{\rho_c}{\alpha c^2}} \right)
\label{eq:mod3phimin}
\end{eqnarray}
where $\phi_{max}$ and $\phi_{min}$ correspond to the positive ($+$) and negative ($-$) signs, respectively. The set of $\alpha$ and $c$ that is in good agreement with the Planck 2018 results \cite{Planck2018} is,
\begin{eqnarray}
\alpha &=& 0.5 m_{Pl}^2, \qquad~~ c = 3.915 \times 10^{-7} m_{Pl}
\label{eq:mod3alphac}
\end{eqnarray}
Other sets of ($\alpha, c$)  that also satisfy the Planck 2018 data are found to yield similar results.  Then, we numerically evolve Eqs. (\ref{eq:Hub}) and (\ref{eq:ddphi}) with (\ref{eq:pot3}) for PIV. Due to the symmetric behavior of the potential, the initial conditions at the bounce have the symmetry $(\phi_B,\dot{\phi}_B) \rightarrow (-\phi_B,-\dot{\phi}_B)$, and the results for NIV can be easily found by applying this symmetry. Furthermore, the initial conditions at the bounce are divided into two sub-cases; KED and PED, and are given by
\begin{equation}
\frac{\phi_B}{m_{Pl}}  =
\begin{cases}
\in  (\phi_{min}, -25.98), & \text{PED (slow-roll)},\cr  
= \pm 25.97, & \text{KE=PE (slow-roll)}, \cr
\in  (-25.96, 25.96), &   \text{KED (slow-roll)},\cr
\in (25.98, \phi_{max}), &  \text{PED (slow-roll)}, \cr
\end{cases}
\label{eq:mod3phiB}
\end{equation}
where $\phi_{max,\; min}$ are given by Eq. (\ref{eq:mod3phimin}). The numerical results for model 3 are presented in Fig. \ref{fig:mod3} with a set of KED and PED initial values at the bounce. 
One of the important result of model 3 in the case $\alpha = 0.5 m_{Pl}^2$ and $c = 3.915 \times 10^{-7} m_{Pl}$ is that we don't get non-slow-roll phase in the entire range of the inflaton field, see Fig. \ref{fig:mod3} and Eq. (\ref{eq:mod3phiB}). However, some of the initial conditions of $\phi_B$ provide less than 60 $e$-folds as shown in table \ref{tab:mod3}, where different inflationary parameters are presented. 
From table \ref{tab:mod3}, one also concludes that $N_{inf}$ grows as the value of $|{\phi}_B|$ increases. Thus, to get enough $e$-folds during the desired slow-roll inflation, the range of $\phi_B$ is restricted to (see table \ref{tab:mod3}),
\begin{eqnarray}
\frac{\phi_B}{m_{Pl}}  =\begin{cases}
\in  (\phi_{min}, -8.22), & N_{inf} \gtrsim 60, \cr
 -8.22 < \frac{\phi_B}{m_{Pl}}  <3.75, & N_{inf} < 60, \cr
 \in   (3.75, \phi_{max}), & N_{inf} \gtrsim 60, \cr
 \end{cases}
\label{eq:mod3phiB60}
\end{eqnarray}
where $\phi_{max, \; min}$ are given by Eq. (\ref{eq:mod3phimin}). 

As mentioned previously, we also numerically studied   other sets of ($\alpha, c$) that satisfy the Planck 2018 data, and found that they give the same results. Therefore, we shall not repeat the calculations again for these cases.

\begin{figure}[tbp]
\begin{center}
\begin{tabular}{ccc}
{\includegraphics[width=1.9in,height=1.65in,angle=0]{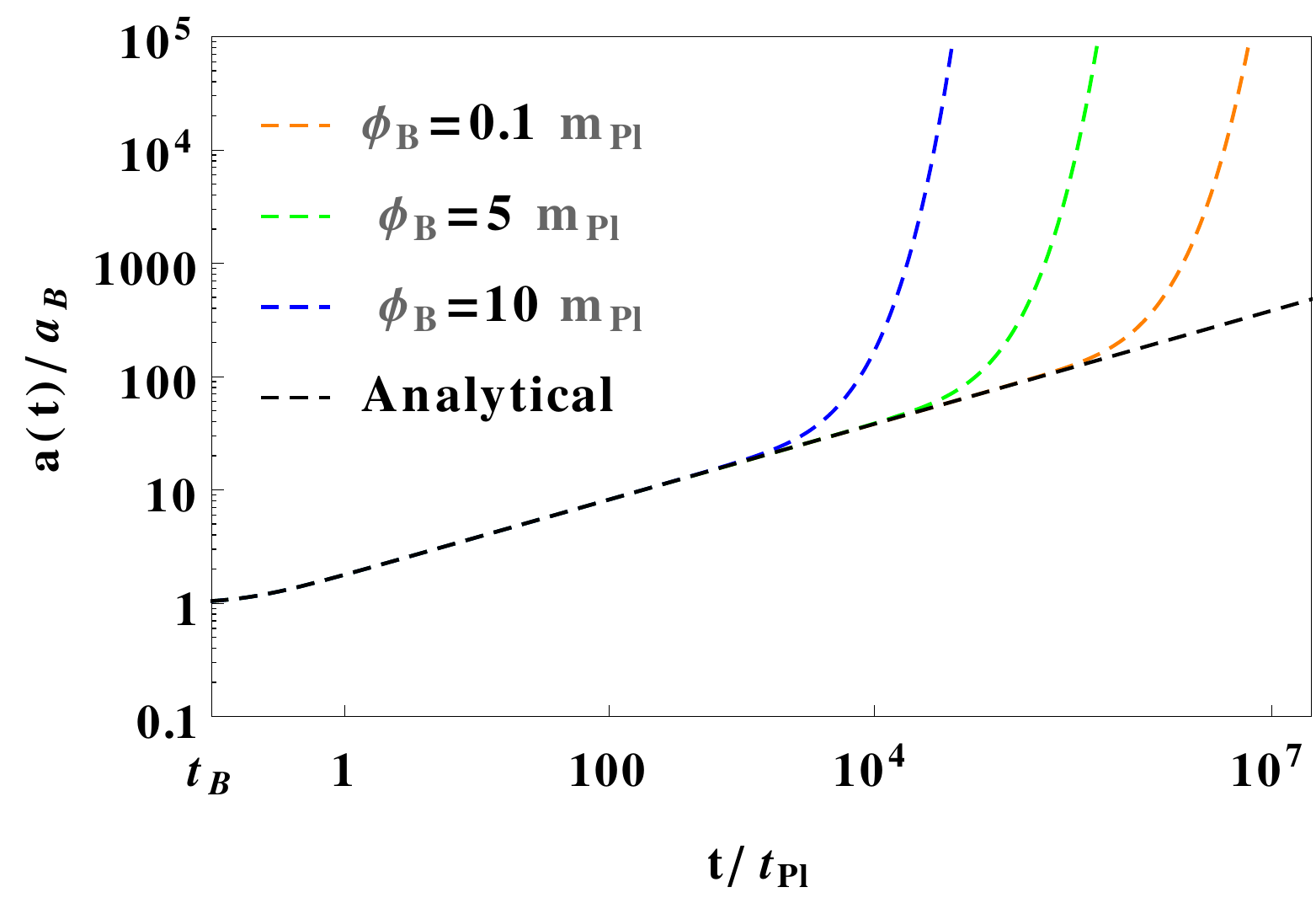}} &
{\includegraphics[width=1.9in,height=1.6in,angle=0]{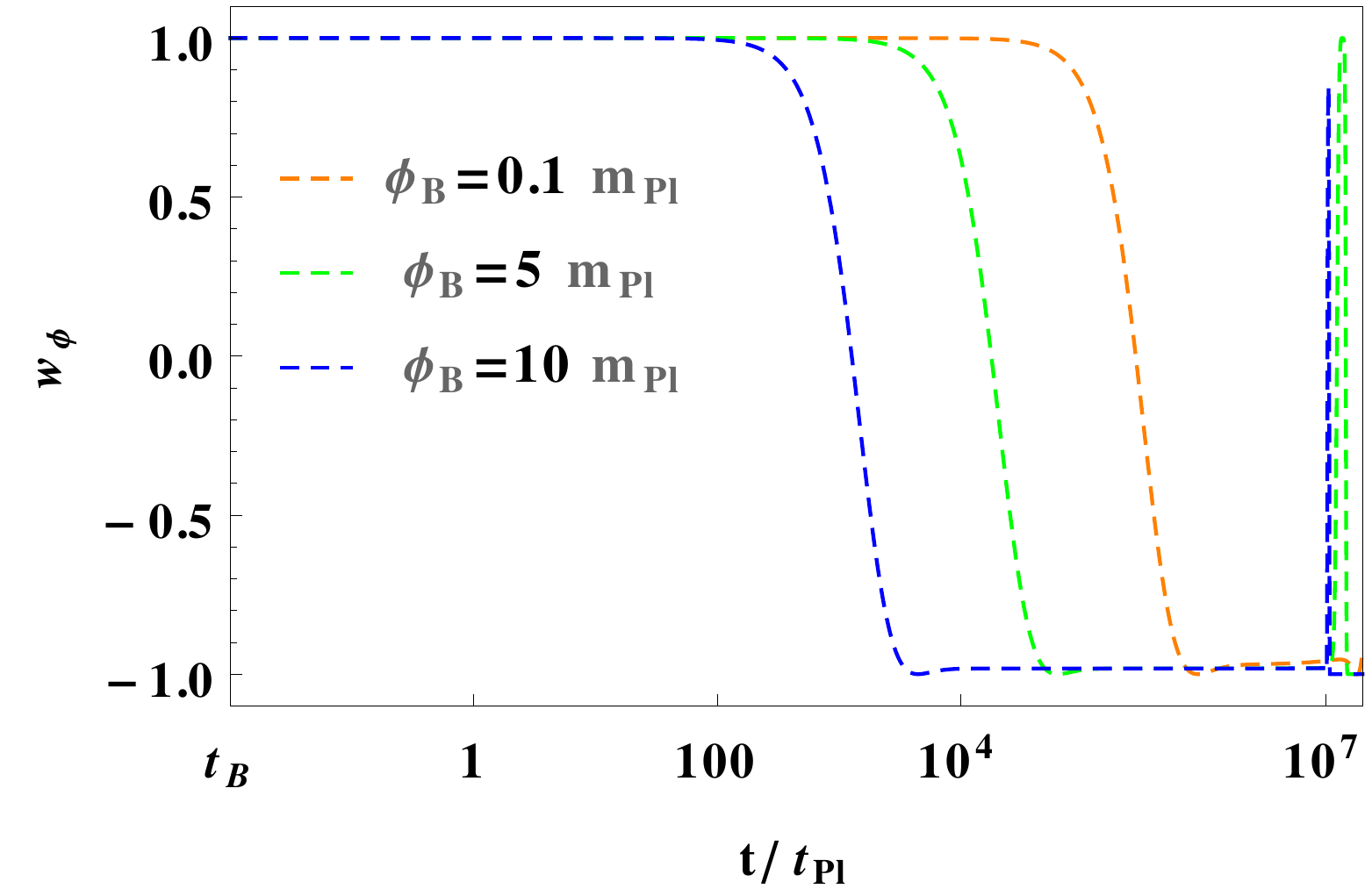}} &
{\includegraphics[width=1.9in,height=1.6in,angle=0]{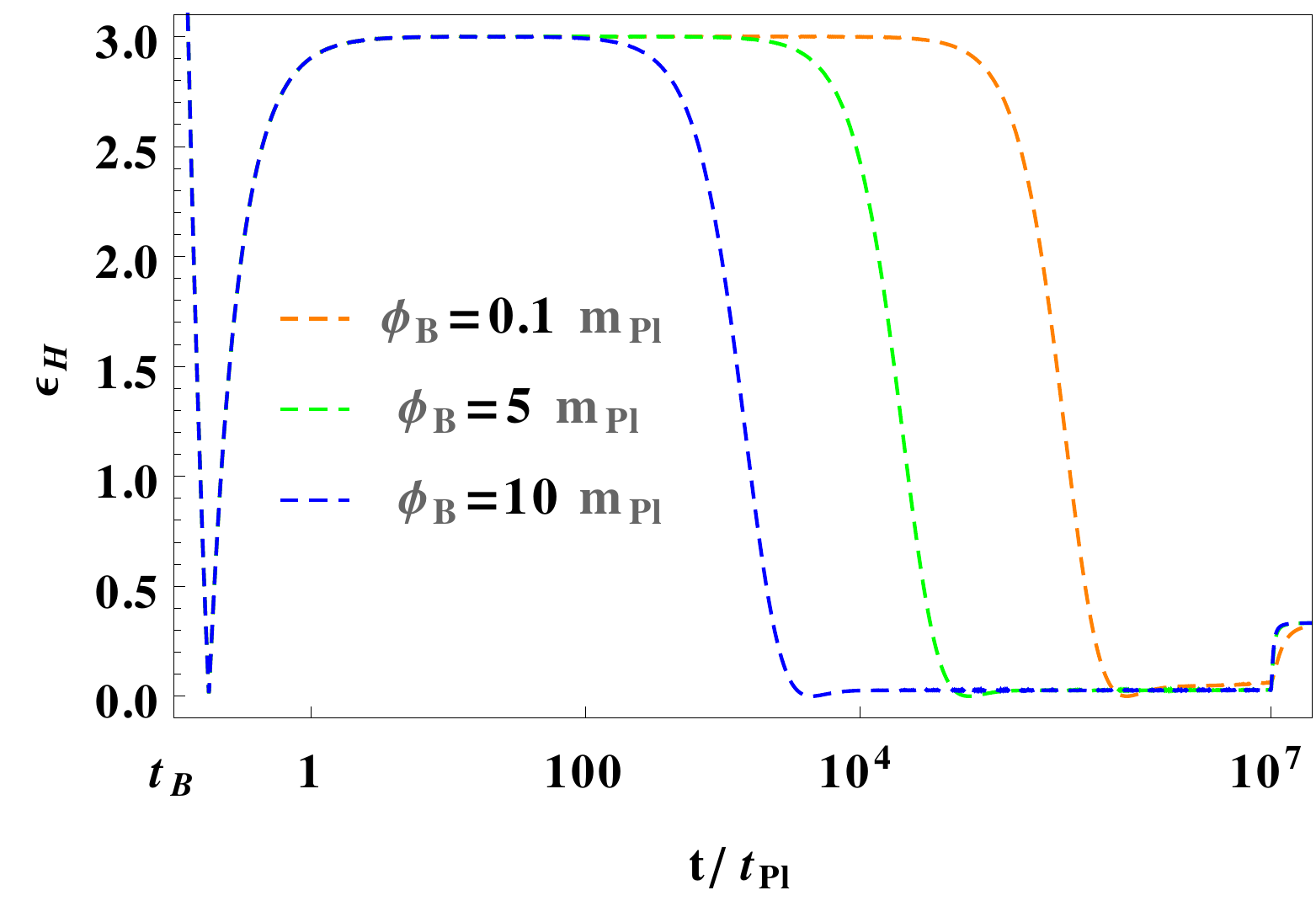}}
\\
{\includegraphics[width=1.9in,height=1.65in,angle=0]{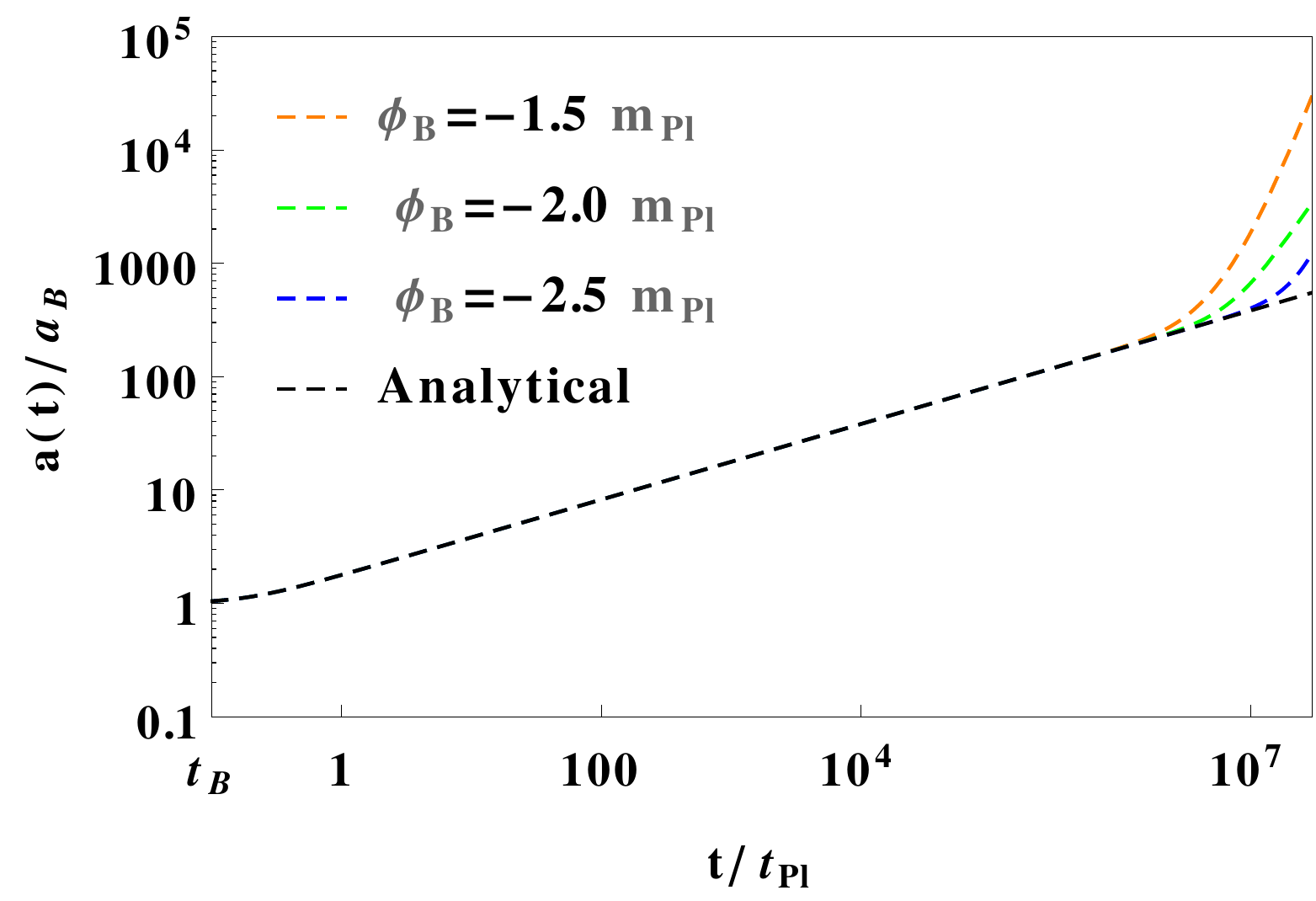}} &
{\includegraphics[width=1.9in,height=1.6in,angle=0]{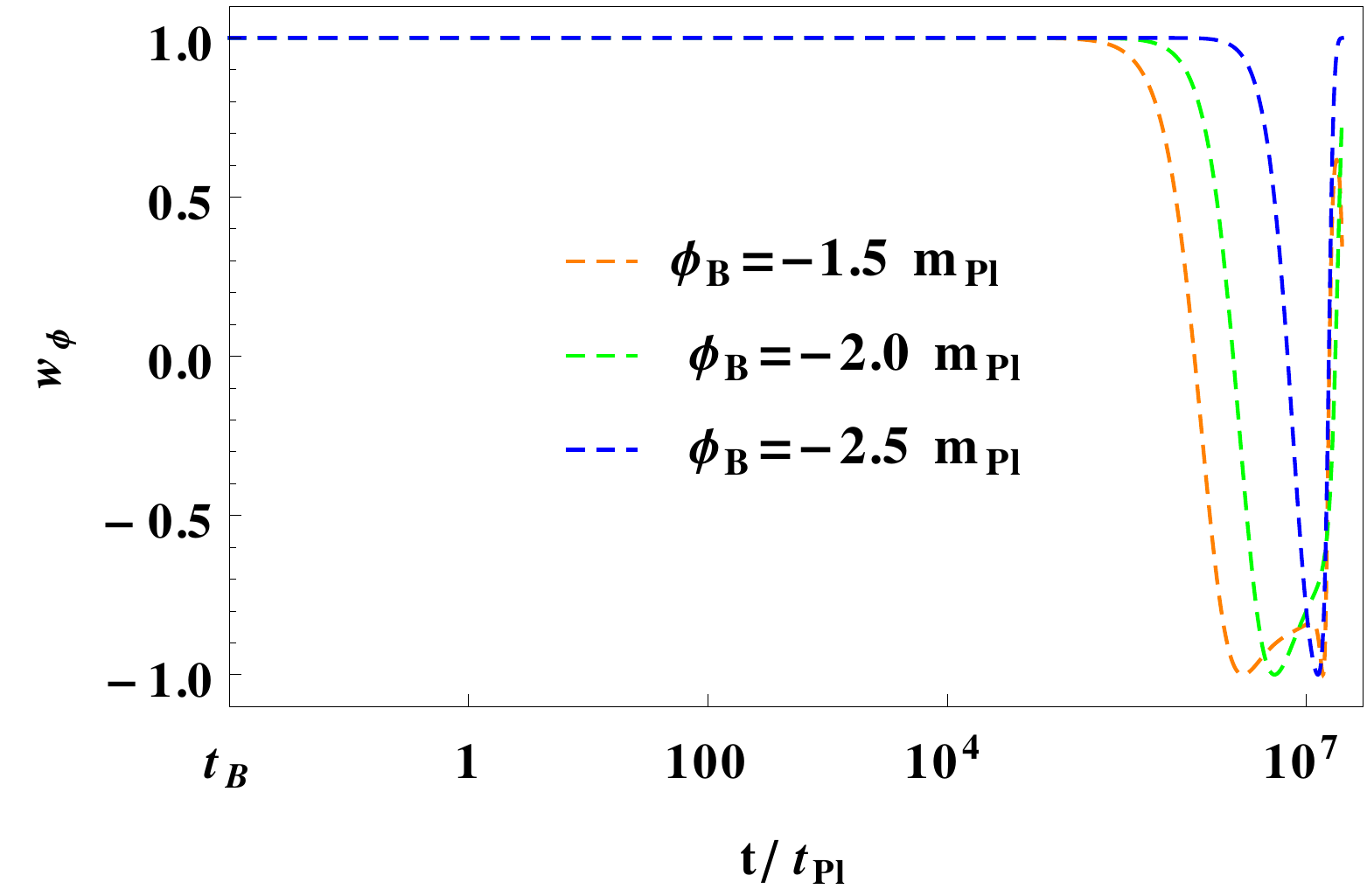}} &
{\includegraphics[width=1.9in,height=1.6in,angle=0]{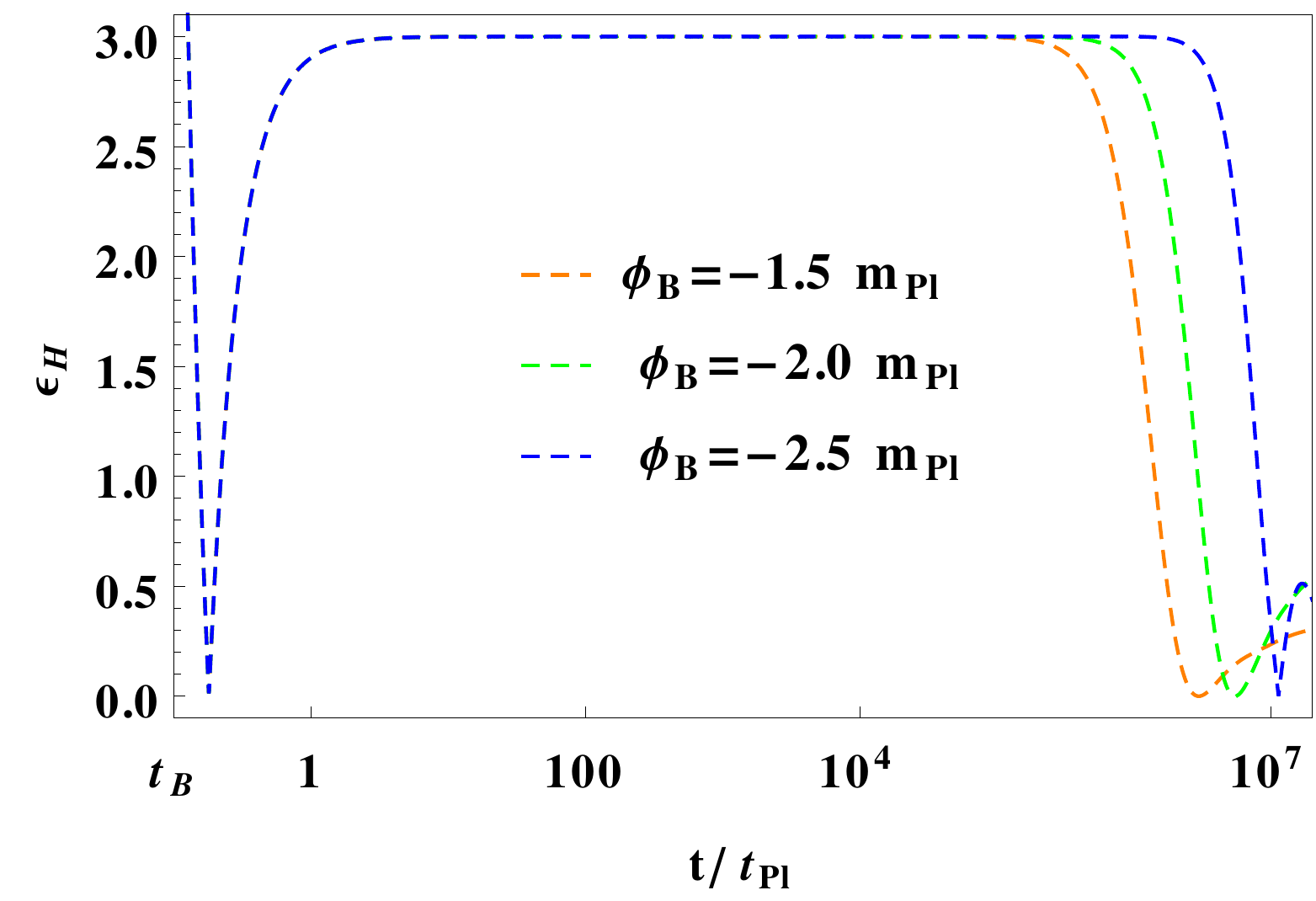}}
\\
{\includegraphics[width=1.9in,height=1.6in,angle=0]{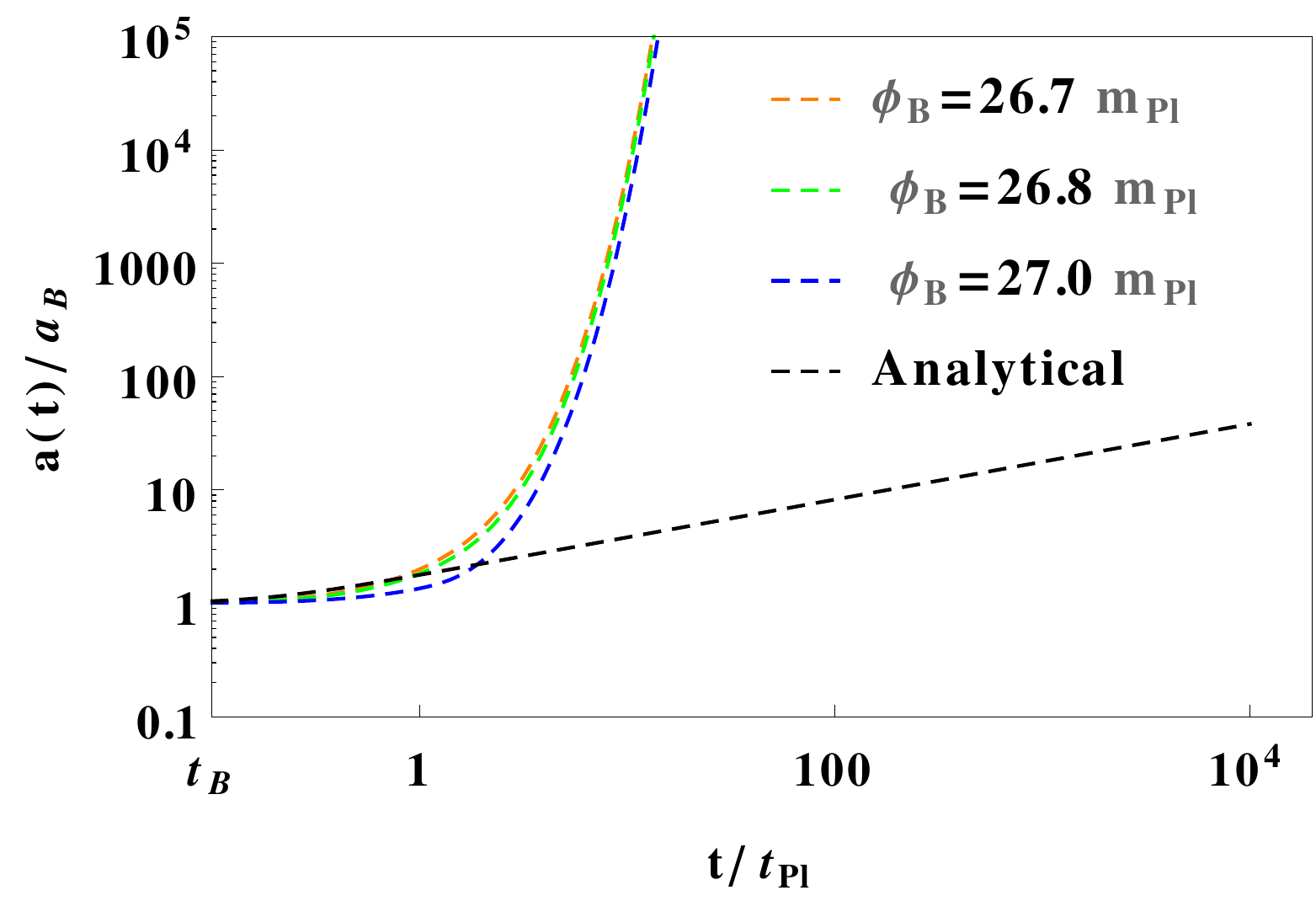}} & 
{\includegraphics[width=1.9in,height=1.6in,angle=0]{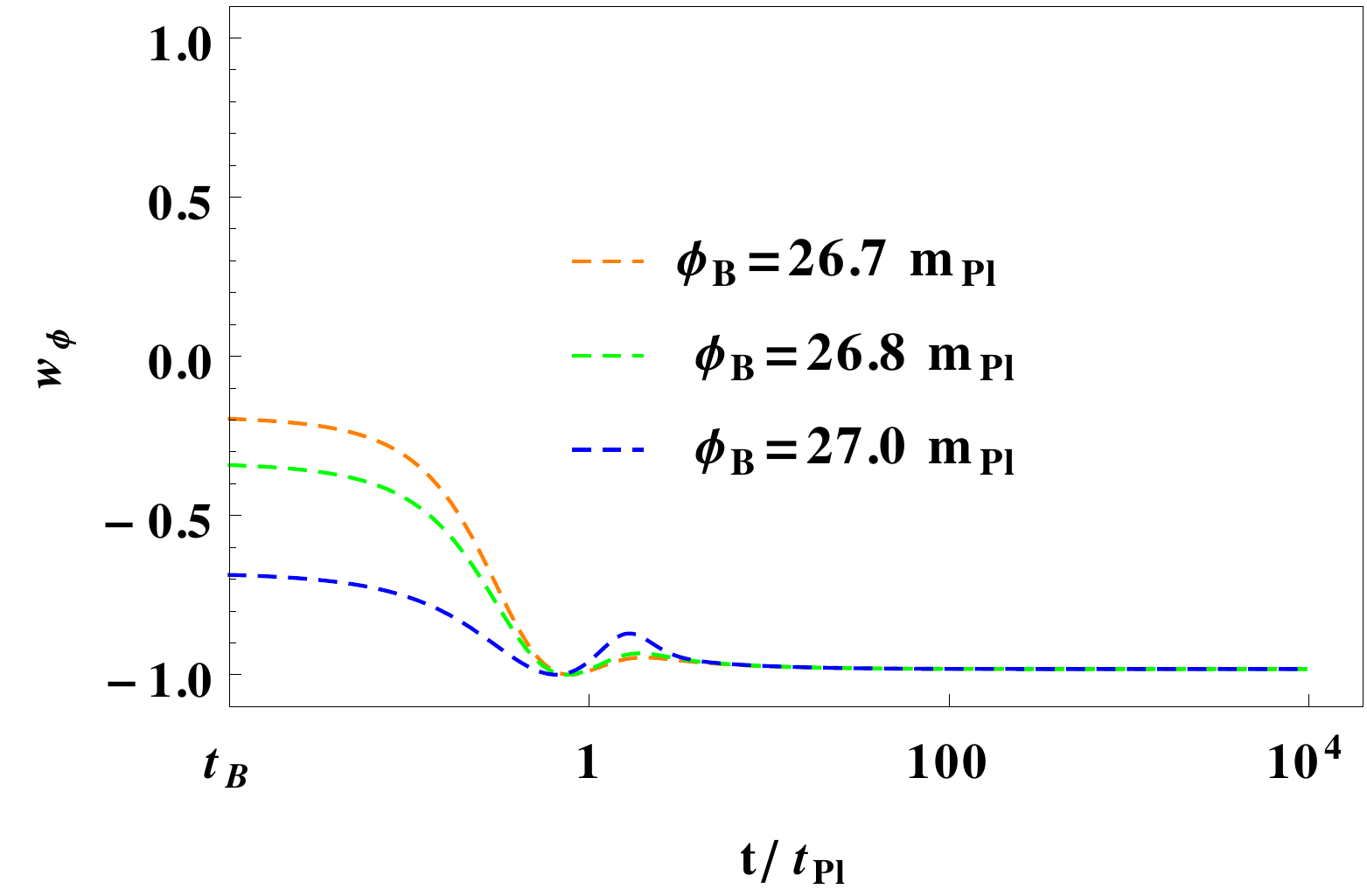}} & 
{\includegraphics[width=1.9in,height=1.6in,angle=0]{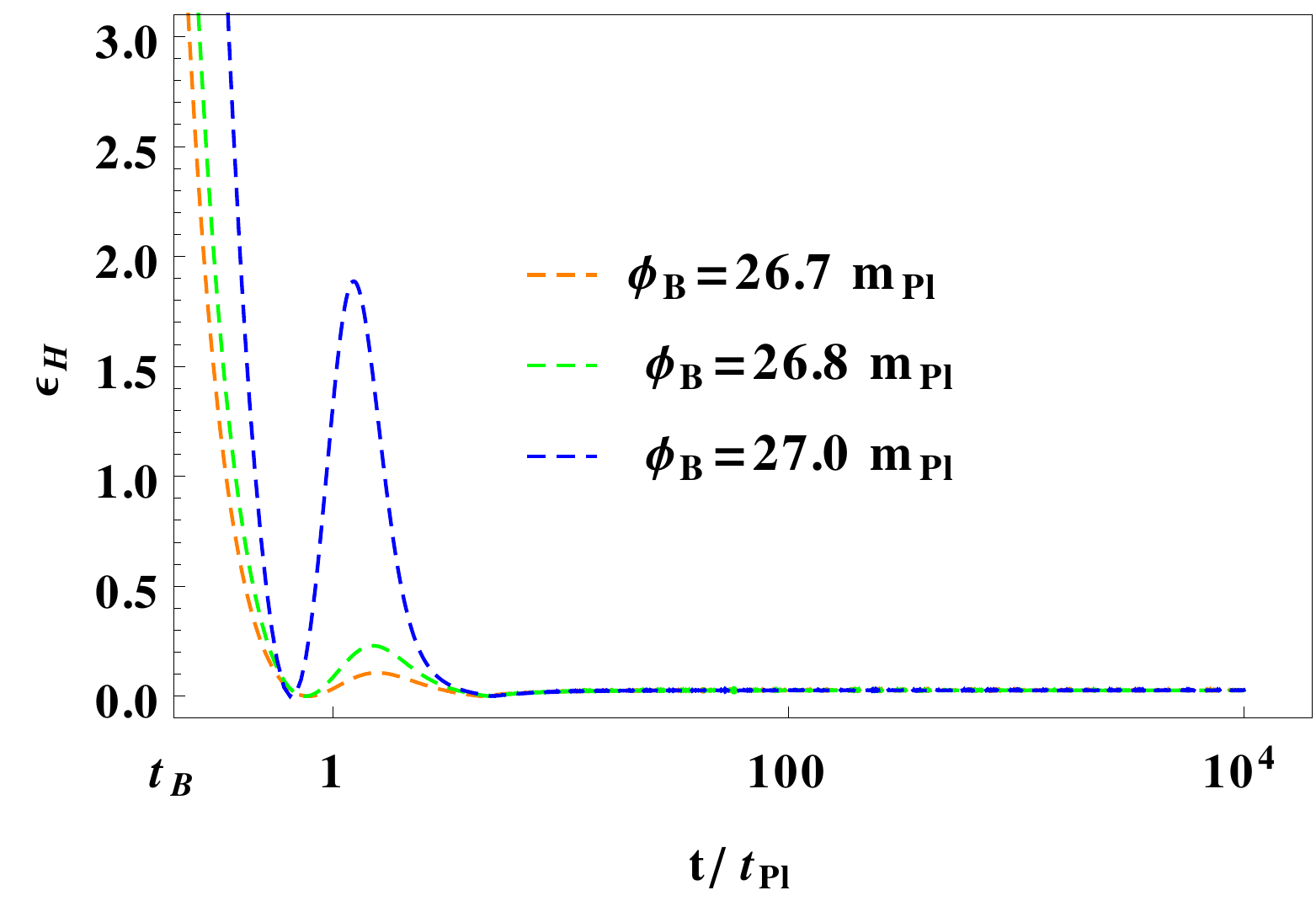}} 
\end{tabular}
\end{center}
\caption{This figure corresponds to model 4 [Eq.(\ref{eq:pot4})] with $\dot{\phi_B}>0$. Due to the symmetric nature of the potential (\ref{eq:pot4}), similar results can be obtained for $\dot{\phi_B}<0$. 
 When plotting out the figure,  we had set $\alpha = 0.5 m_{Pl}^2$ and  $c = 2.818 \times 10^{-7} m_{Pl}$ and $m_{Pl}=1$.  }
\label{fig:mod4}
\end{figure}
\begin{table}[tbp]
\caption{This table is displayed for model 4 [Eq.(\ref{eq:pot4})] with $\dot{\phi}_B > 0$, and $\alpha = 0.5 m_{Pl}^2$ and  $c = 2.818 \times 10^{-7} m_{Pl}$.}
\begin{center}
\resizebox{\textwidth}{!}{
\begin{scriptsize}
\begin{tabular}{cccccccccc}
\hline
$\phi_B/m_{Pl}$~~~  & Inflation~~~ & $t/t_{Pl}$~~~ & $\epsilon$~~ & $w$ ~~& $N_{inf}$ &~~~${w}^B$\\
\hline
$26.7$ ~~~& begin~~~& 0.11 ~~~& 4.5~~ & $-1/3$ ~~& ~~~& ~~~&\\
& slow-roll~~~& 1.22 ~~~& 0.080~~ & $-0.970$ ~~& 479.76 ~~~& $<0$\\
& end~~~& $1.25508 \times 10^7$ ~~~& 0.326~~ & $-1/3$ ~~& ~~~& ~~~& \\\\
5 ~~~& begin~~~& 2.25906 $\times 10^4$~~~& 0.999~~ & $-1/3$ ~~& ~~~& ~~~&\\
& slow-roll~~~& 6.25623$\times 10^4$ ~~~& 3.04$\times 10^{-5}$~~ & $-1$ ~~& 69.27 ~~~& $>0$\\
& end~~~& $1.21626 \times 10^7$ ~~~& 0.318~~ & $-1/3$ ~~& ~~~& ~~~& \\\\
4 ~~~& begin~~~& 3.83533 $\times 10^4$~~~& 0.999~~ & $-1/3$ ~~& ~~~& ~~~&\\
& slow-roll~~~& 9.9128$\times 10^4$ ~~~& 1.49$\times 10^{-3}$~~ & $-0.999$ ~~& 60.65 ~~~& $>0$\\
& end~~~& $3.18112 \times 10^7$ ~~~& 0.333~~ & $-1/3$ ~~& ~~~& ~~~& \\\\
3.5 ~~~& begin~~~& 5.00098 $\times 10^4$~~~& 0.999~~ & $-1/3$ ~~& ~~~& ~~~&\\
& slow-roll~~~& 1.14321$\times 10^5$ ~~~& 1.50$\times 10^{-2}$~~ & $-0.990$ ~~& 46.73 ~~~& $>0$\\
& end~~~& $1.30343 \times 10^7$ ~~~& 0.322~~ & $-1/3$ ~~& ~~~& ~~~& \\\\
$-8$ ~~~& begin~~~& 4.61463 $\times 10^4$~~~& 1.0~~ & $-1/3$ ~~& ~~~& ~~~&\\
& slow-roll~~~& 2.40478$\times 10^5$ ~~~& 2.99$\times 10^{-2}$~~ & $-0.980$ ~~& 45.14 ~~~&$>0$\\
& end~~~& $1.12755 \times 10^7$ ~~~& 0.285~~ & $-1/3$ ~~& ~~~& ~~~& \\\\
$-8.73$ ~~~& begin~~~& 2.89127 $\times 10^4$~~~& 1.0~~ & $-1/3$ ~~& ~~~& ~~~&\\
& slow-roll~~~& 1.48336$\times 10^5$ ~~~& 2.99$\times 10^{-2}$~~ & $-0.979$ ~~& 60.61 ~~~&$>0$\\
& end~~~& $1.32331 \times 10^7$ ~~~& 0.325~~ & $-1/3$ ~~& ~~~& ~~~& \\\\
$-9$ ~~~& begin~~~& 2.43324 $\times 10^4$~~~& 1.0~~ & $-1/3$ ~~& ~~~& ~~~&\\
& slow-roll~~~& 1.24506$\times 10^5$ ~~~& 2.99$\times 10^{-2}$~~ & $-0.980$ ~~& 64.18 ~~~& $>0$\\
& end~~~& $1.16183 \times 10^7$ ~~~& 0.308~~ & $-1/3$ ~~& ~~~& ~~~& \\
\hline
\end{tabular}
\end{scriptsize}}
\label{tab:mod4}
\end{center}
\end{table}

\subsection{Model 4}
\label{subsec:mod4}

Finally, we consider the case with the  potential (\ref{eq:pot4}) (model 4). The evolution of this potential is shown in the lower right panel of Fig. \ref{fig:pot}. The potential is bounded below by zero ($V(\phi) \geq 0$) and unbounded from above, and oscillates around the origin ($\phi = 0$). The behavior of this potential is symmetric with respect to $\phi=0$. In the large field limit ($\phi \rightarrow \pm \infty$), the critical energy density $\rho_c$ constrains the initial conditions of the inflaton field at the bounce that depends on the value of $\alpha$ and $c$. The following combination of $\alpha$ and $c$ is compatible with the Planck 2018 data \cite{Planck2018} (see appendix)
\begin{eqnarray}
\alpha &=& 0.5 m_{Pl}^2, \qquad~~ c = 2.818 \times 10^{-7} m_{Pl}.
\label{eq:mod4alphac}
\end{eqnarray}
In this subsection, we shall investigate the dynamics of the pre-inflationary universe with 
such given $\alpha$ and $c$ only for $\dot\phi_B >0$, and the other possibilities ($\dot\phi_B <0$, as well as in  other sets
of $\alpha$ and $c$) will yield similar results. The corresponding value of $\phi_{max,\; min}$ at the bounce will be $\pm27.2 m_{Pl}$. Similar to model 3,  in model 4 the potential is also symmetric. Therefore, we shall not consider the NIV case,  due to the symmetry $(\phi_B,\dot{\phi}_B) \rightarrow (-\phi_B,-\dot{\phi}_B)$. We numerically solve Eqs. (\ref{eq:Hub}) and (\ref{eq:ddphi}) with potential (\ref{eq:pot4}) for $\alpha = 0.5 m_{Pl}^2$ and $c = 2.818 \times 10^{-7} m_{Pl}$. The results are illustrated in Fig. \ref{fig:mod4}. we obtain a subset of initial conditions that does not provide the slow-roll inflation as shown in the middle panel of Fig. \ref{fig:mod4}. The rest of the cases (KED \& PED) will be quite similar to those studied  in model 3, so we shall not repeat the analysis here, but simply summarize  the final results with various ranges of  the  initial conditions of $\phi_B$,
\begin{eqnarray}
\frac{\phi_B}{m_{Pl}} = \begin{cases}
 \in  (-27.2, -26.57), & \text{PED (slow-roll)},\cr
= \pm 26.56, & \text{KE=PE (slow-roll)}, \cr
\in  (-26.55, -5.1), & \text{KED (slow-roll)},\cr
 -5 \leqslant \frac{\phi_B}{m_{Pl}} < -0.1, & \text{KED (no slow-roll)},\cr
\in  (-0.1, 26.55), & \text{KED (slow-roll)},\cr
\in (26.57, 27.2), & \text{PED (slow-roll)}.\cr
\end{cases}
\label{eq:mod4phiB}
\end{eqnarray}
The results of model 4 are shown in Fig. \ref{fig:mod4} and table \ref{tab:mod4}. Again, we shall not explain   the detail of Fig. \ref{fig:mod4},  as the evolution is quite similar to model 3. However, we obtain a subset of initial conditions that does not provide the slow-roll phase. By looking at table \ref{tab:mod4}, the physical viable initial conditions of $\phi_B$ that generate enough $e$-folds for the desired slow-roll inflation are
\begin{eqnarray}
\frac{\phi_B}{m_{Pl}}  = \begin{cases}
\in  (4, 27.2), & N_{inf} \gtrsim 60, \cr 
 \in   (-8.73, -27.2), &N_{inf} \gtrsim 60.\cr
\end{cases}
\label{eq:mod4phiB60}
\end{eqnarray}
Within these ranges, $N_{inf}$ always increases as  $|{\phi}_B|$ grows.

\section{Phase portraits and desired slow-roll inflation}
\label{sec:port}

Let us investigate the phase spaces for the models under our considerations. First, we consider  model 1 for $\alpha = 1 m_{Pl}^2$ and  $c = 8.343 \times 10^{-7} m_{Pl}$.
In this case, as shown previously, the entire range of the initial conditions   does not yield  a desired slow-roll inflation with enough $e$-folds, which  are  inconsistent with the observational data, as shown explicitly in table \ref{tab:mod1}. Hence, we shall not draw the phase portrait for model 1.

\begin{figure}[tbp]
\begin{center}
\begin{tabular}{ccc}
{\includegraphics[width=1.9in,height=2in,angle=0]{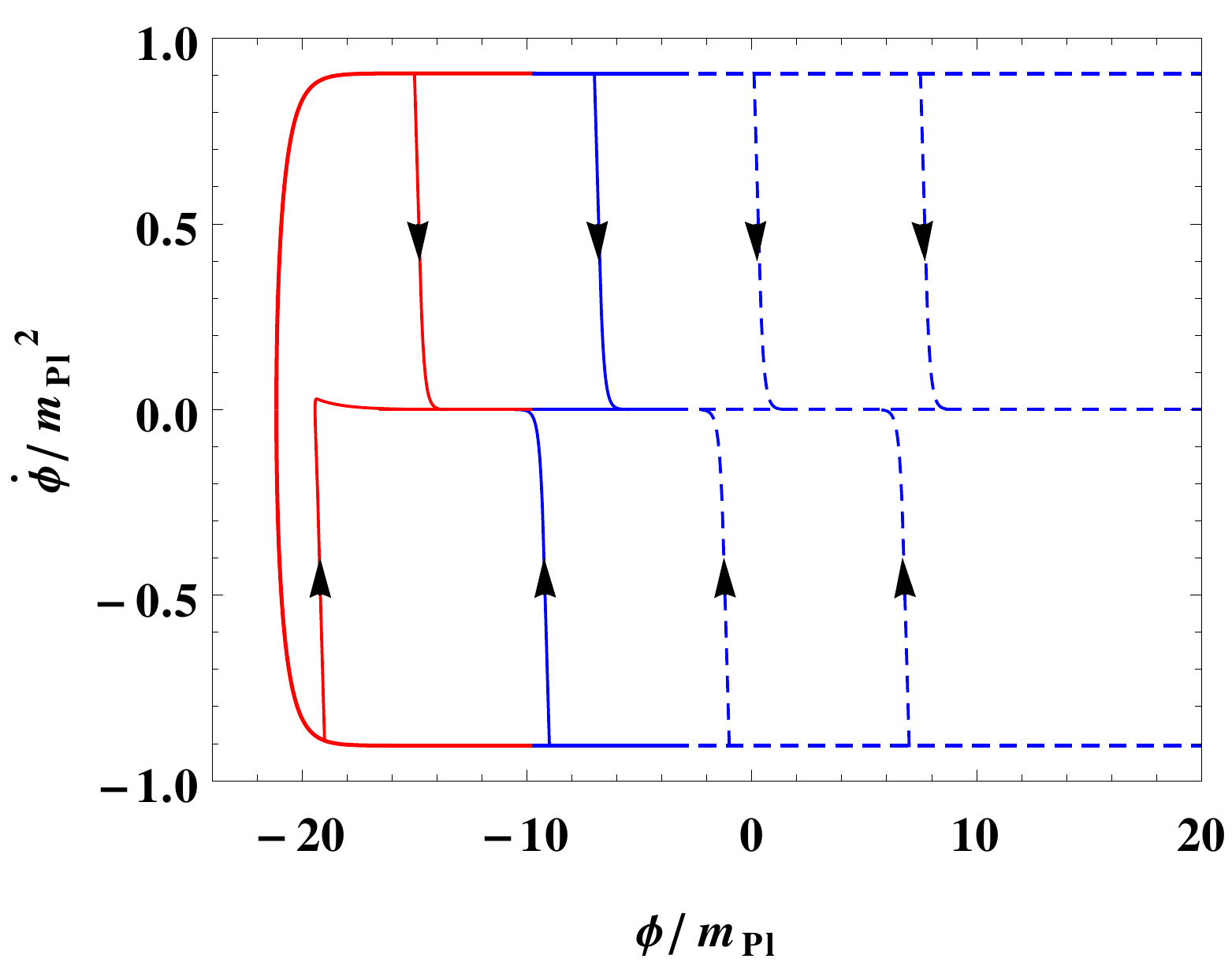}} &
{\includegraphics[width=1.9in,height=2in,angle=0]{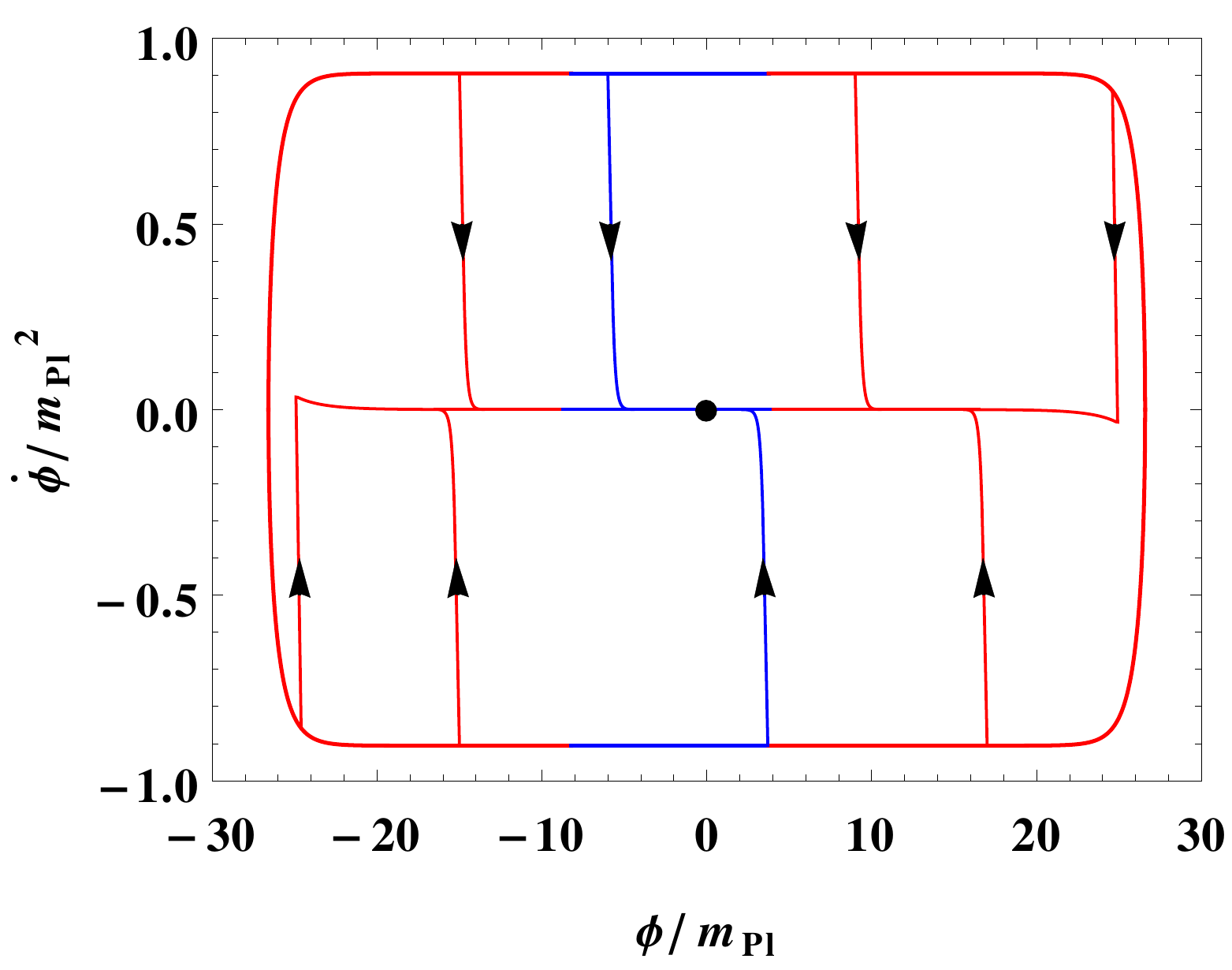}} &
{\includegraphics[width=1.9in,height=2in,angle=0]{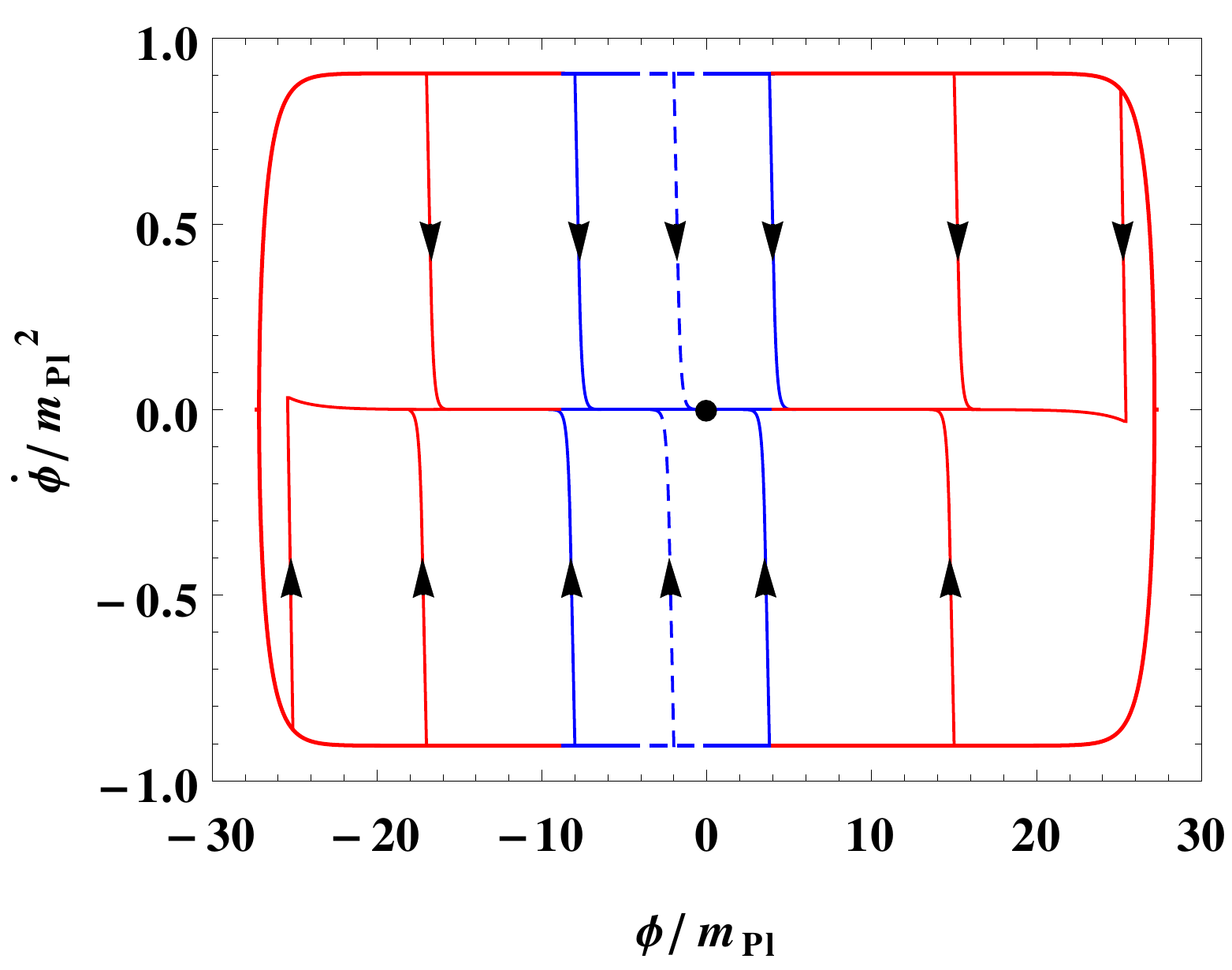}}
\end{tabular}
\end{center}
\caption{This figure shows the phase portraits of models 2 (left), 3 (middle) and 4 (right) in the $(\phi/m_{Pl}, \dot{\phi}/m_{Pl}^2)$ plane. All trajectories (with arrowheads) start at the bounce at which we have $\rho=\rho_c$ (boundary surface without arrowheads). The red trajectories generate the desired slow-roll inflation, while the blue (solid) ones do not. The dashed (blue) trajectories demonstrate the case without inflation. In model 2 (left; $\alpha = 1 m_{Pl}^2$ and  $c = 4.074 \times 10^{-8} m_{Pl}$), the initial data is in the range,    $\phi/m_{Pl} \in (\phi_{min}, \infty)$ (see Eq.(\ref{eq:mod2phiB})), but here we show only a part of it. Since the left panel extends from $\phi_{min}$ to $\infty$, the length of the blue curves (solid and dashed) is very long in comparison with the red ones. Therefore, a slow-roll inflation exists for a short period. For models 3 and 4, the initial surface extends to $\phi/m_{Pl} \rightarrow \pm 26.58$ (middle panel; $\alpha = 0.5 m_{Pl}^2$ and  $c = 3.915 \times 10^{-7} m_{Pl}$) and $\phi/m_{Pl} \rightarrow \pm 27.2$ (right panel; $\alpha = 0.5 m_{Pl}^2$ and  $c = 2.818 \times 10^{-7} m_{Pl}$), respectively. In the middle and right panels, the lengths of the blue trajectories are very short   in comparison with the red ones. As a result,  the slow-roll inflation is almost inevitable.  }
\label{fig:port}
\end{figure}

Second, we examine the phase portrait for model 2 with  $\alpha = 1 m_{Pl}^2$ and  $c = 4.074 \times 10^{-8} m_{Pl}$. In this case, we find the inflationary and non-inflationary phases for different sets of $\phi_B$ as displayed in Figs. \ref{fig:mod2} and \ref{fig:port}. The left panel of Fig. \ref{fig:port} exhibits the evolution of the phase space trajectories in the $(\phi/m_{Pl}, \dot{\phi}/m_{Pl}^2)$ plane for both of the PIV and NIV cases, and also for the KED and PED initial conditions. The initial data surface is semi-finite: $| \dot{\phi}_B |/m_{Pl}^2  < 0.91 $ and $\phi_B/m_{Pl} \in (-21.14, \infty)$ due to the shape of the potential (\ref{eq:pot2}). The solid (blue) trajectories correspond to the inflationary region that do not provide the desired slow-roll inflation as the number of $e$-folds is not sufficient. The dashed (blue) trajectories exhibit the non-inflationary region. Only the red trajectories demonstrate the desired slow-roll inflation that are consistent with observations, that is, a slow-roll inflationary phase with enough e-folds. Likewise, the solid and dashed (blue) parts of the boundary surface is governed by the inflationary (not consistent with observations as it does not generate sufficient $e$-folds) and non-inflationary phases,  while the red surface is in good agreement with observations as it produces at least  60 $e$-folds and more. From Eqs. (\ref{eq:mod2phiB}), (\ref{eq:mod2phiB60}) and the left panel of Fig. \ref{fig:port}, one can see that the region of the desired slow-roll inflation is less than the region of the non-inflationary phase, and also less than the part that does not give the desired slow-roll inflation. Hence, in this case only a small portion  of the initial conditions produce the desired slow-roll inflation with sufficient e-folds. In the left panel of Fig. \ref{fig:port}, we show this small portion of the initial conditions,  while the whole range is given by Eq. (\ref{eq:mod2phiB}). 

Next, we carry out the phase space analysis for model 3 with $\alpha = 0.5 m_{Pl}^2$ and $c = 3.915 \times 10^{-7} m_{Pl}$. The phase portrait for this model  is depicted in the middle panel of Fig. \ref{fig:port}. The initial data surface is totally compact: $| \dot{\phi}_B |/m_{Pl}^2 < 0.91  $ and $\phi_B/m_{Pl} \rightarrow \pm 26.58$, as the critical energy density $\rho_c$ puts the bound on the initial values of $\phi_B$. The red trajectories and surface generate the desired slow-roll inflation which is compatible with observations,  whereas the blue ones are not. The middle panel of Fig. \ref{fig:port} exhibits the evolution of PIV and NIV, and also for the KED and PED initial values at the bounce. More preciously, it covers the whole phase space. Regions close to the boundary correspond to the large energy density where the quantum effects dominate,  while the low energy limit exists near the origin in the $(\phi/m_{Pl}, \dot{\phi}/m_{Pl}^2)$ plane. All curves start from the surface of the bounce ($\rho=\rho_c$) and move  towards the origin which is a single stable point. In the entire phase space, the blue region is much less than the red one. Therefore,   in this model a substantial fraction of initial values of the inflaton field produces the desired slow-roll inflation, and  the occurrence of a slow-roll inflation is practically inevitable.

Finally,   for model 4 with $\alpha = 0.5 m_{Pl}^2$ and $c = 2.818 \times 10^{-7} m_{Pl}$, the phase portrait is presented in the right panel of  Fig. \ref{fig:port}. In model 4, the boundary surface is also finite: $| \dot{\phi}_B |/m_{Pl}^2 < 0.91  $ and $\phi_B/m_{Pl} \rightarrow \pm 27.2$. In this case, we get non-inflationary phases. The rest of the analysis is quite similar to model 3, so we shall not repeat it. 

\section{Conclusions}
\label{sec:conc}

In this paper, we studied the dynamics of the pre-inflationary universe with a family of $\alpha-$attractor potentials for $\dot{\phi}_B > 0$ in the framework of LQC. First, we investigated numerically  the background evolution for model 1 with $\alpha = 1 m_{Pl}^2$ and  $c = 8.343 \times 10^{-7} m_{Pl}$. 
In this case, the initial conditions at the bounce are dominated only by KE as the PED initial conditions do not exist during the whole bouncing phase. Similar results were obtained for $T-model$ in Ref. \cite{alamPRD2018}.
The numerical results for model 1 are presented in Fig. \ref{fig:mod1}, where $a(t)$, $w(\phi)$ and $\epsilon_H$ are displayed for several values of  ${\phi}_B$. From the numerical evolution of $w(\phi)$, one can see that the universe is split  into three different phases prior to reheating: {\it bouncing, transition and the slow-roll inflation}. During the bouncing phase, the evolution of $a(t)$ is universal  for a wide range of initial conditions, and is well described  by the analytical solution (\ref{eq:a}), as shown in the left panel of Fig. \ref{fig:mod1}. In this phase, $w(\phi) \simeq +1$. However, it decreases quickly from $w(\phi) \simeq +1$  to $w(\phi) \simeq -1$ during the transition phase, and then stays pegged at  $w(\phi) \simeq -1$ in the slow-roll phase. The period of transition phase is very short in comparison with the other two phases. We also found the number of $e$-folds during the slow-roll inflation that is shown in Table \ref{tab:mod1}. For model 1, we always get less than 60 $e$-folds during the slow-roll inflationary phase for any given value of   ${\phi}_B$ in the range. Hence, this model is not observationally favorable. 

Second, we studied numerically the evolution of the background for model 2 with $\alpha = 1 m_{Pl}^2$ and  $c = 4.074 \times 10^{-8} m_{Pl}$. In the case of $\alpha = 1 m_{Pl}^2$ and  $c = 4.074 \times 10^{-8} m_{Pl}$, the range of ${\phi}_B$ is divided into the KED and PED initial conditions, and the numerical results are presented in Fig. \ref{fig:mod2}. For the KED  case (except for a subset), the evolution of the scale factor $a(t)$ during the bouncing phase shows universal feature, that is, it does not depends on initial conditions and is well described  by the  analytical solution (\ref{eq:a}). During the bouncing phase, the EoS $w(\phi) \simeq +1$. It drastically decreases from $+1$ to $-1$ in the transition phase. Soon, the universe enters into  the slow-roll phase, where $\epsilon_H$ is still large initially, but quickly declines to zero, and the slow-roll inflation takes place, as shown by the upper panels of Fig. \ref{fig:mod2}. A subset of the KED initial conditions does not lead to inflation as shown in the middle panels of Fig. \ref{fig:mod2}. In the case of the PED initial conditions, the universality of $a(t)$ is lost. Bouncing and transition phases do not exist any more. Though, the slow-roll inflation can still be achieved for a long period. We also showed other parameters in Table \ref{tab:mod2}, where  physically viable initial conditions of ${\phi}_B$  were identified, which produce  enough  $e$-folds. From Table \ref{tab:mod2}, we can see that $N_{inf}$ decreases as ${\phi}_B$ grows.

On the other hand, for models 3 and 4, we examined numerically the background evolutions with  $\alpha = 0.5 m_{Pl}^2$ and $c = 3.915 \times 10^{-7} m_{Pl}$ (model 3) and $\alpha = 0.5 m_{Pl}^2$ and $c = 2.818 \times 10^{-7} m_{Pl}$ (model 4), respectively. The results are shown in Figs. \ref{fig:mod3} and \ref{fig:mod4}. The whole range of the initial values of  ${\phi}_B$ provide the slow-roll inflationary phase for model 3,  whereas in  model 4, a subset of the initial conditions exists without inflation. 
The number of $e$-folds $N_{inf}$ and other inflationary parameters are displayed in Tables \ref{tab:mod3} and \ref{tab:mod4}, where $N_{inf}$ increases as the absolute value of ${\phi}_B$ grows.

Finally, we presented the phase portraits for models 2, 3 and 4 in Fig. \ref{fig:port}. We did not display the phase portrait for model 1 as all the initial conditions of inflaton field provide less than 60 $e$-folds that are not consistent with observations. For model 2 with $\alpha = 1 m_{Pl}^2$ and  $c = 4.074 \times 10^{-8} m_{Pl}$, the quantum bounce surface is semi-finite: $| \dot{\phi}_B |/m_{Pl}^2  < 0.91 $ and $\phi_B/m_{Pl} \in (-21.14, \infty)$,  whereas for models 3 and 4, the bounce surface is compact. In particular, in model 3 with
$\alpha = 0.5 m_{Pl}^2$ and $c = 3.915 \times 10^{-7} m_{Pl}$, we found  $| \dot{\phi}_B |/m_{Pl}^2 < 0.91  $ and $\phi_B/m_{Pl} \rightarrow \pm 26.58$,  while for model 4 with $\alpha = 0.5 m_{Pl}^2$ and $c = 2.818 \times 10^{-7} m_{Pl}$, we obtained $| \dot{\phi}_B |/m_{Pl}^2 < 0.91  $ and $\phi_B/m_{Pl} \rightarrow \pm 27.2$. In Fig. \ref{fig:port}, the dashed blue trajectories correspond to the case without inflation,  and the solid trajectories (red and blue) can lead to the slow-roll inflation. However, only the red curves generate sufficient $e$-folds that are compatible with the Planck 2018 data, and not the blue ones \cite{Planck2018}.

\acknowledgments
A.W. would like to thank  ITPC - ZJUT for their hospitality during the summer of 2019, in which part of the work was done. His research is supported in part by the National Natural Science Foundation of China (NNSFC) with the Grants Nos. 11975203 and 11675145. M. Al Ajmi  is supported by Sultan Qaboos University under the Internal Grant (IG/SCI/PHYS/19/02). Part of the work is also supported by  the Ministry of Education and Science,  the Republic of Kazakhstan, with Grant No. 0118RK00693.

\appendix
\section{Some Physical Quantities}
\label{sec:Append}
\renewcommand{\theequation}{A.\arabic{equation}}\setcounter{equation}{0}

From  Eq.(\ref{eq:Ninf}), we have
\begin{eqnarray}
N_{inf} \simeq \int_{\phi_{end}}^{\phi_*} \frac{V(\phi)}{V'({\phi})} d\phi, 
\label{eq:Ninf2}
\end{eqnarray}
where $\phi_*$ and $\phi_{end}$ represent the values of the inflaton field at the beginning and end of the slow-roll inflation.

The slow-roll parameter $\epsilon_V$ is given by
\begin{eqnarray}
\epsilon_V = \frac{M_{Pl}^2}{2} \left(\frac{V'(\phi)}{V(\phi)}\right)^2.
\label{eq:ev}
\end{eqnarray}
At the end of the slow-roll inflation,  $\epsilon_V=1$. Hence, one can find $\phi_{end}$ from  Eq.(\ref{eq:ev}).

During the slow-roll inflation, $\dot{\phi}^2 \ll V(\phi)$. Therefore,  Eq.(\ref{eq:Hub}) becomes
\begin{eqnarray}
H_*{^2} \simeq \frac{8 \pi}{3 m_{Pl}^2}~V(\phi_{*}).
\label{eq:HubSR}
\end{eqnarray}
According to the Planck 2018 results \cite{Planck2018}, the upper bound on $H_*$ during the slow-roll inflation is given by 
\begin{eqnarray}
\frac{H_*}{M_{Pl}} < 2.5 \times 10^{-5} ~~ (\text{95 \% Confidence level}).
\label{eq:H*}
\end{eqnarray}
In our current work, we choose $H_*{/M_{Pl}}=2.0 \times 10^{-5}$. Substituting  the value of $H_*{/M_{Pl}}$ into  Eq.(\ref{eq:HubSR}), we obtain $\phi_*$.
By putting the values of  $\phi_*$ and $\phi_{end}$ with $N_{inf}=60$ in  Eq.(\ref{eq:Ninf2}), we get different combinations of $\alpha$ and $c$, as shown in Eqs. (\ref{eq:mod1alphac}), (\ref{eq:mod2alphac}), (\ref{eq:mod3alphac}) and (\ref{eq:mod4alphac}).



\begin{thebibliography}{99}
\bibitem{guth1981} A. H. Guth, Inflationary universe: A possible solution to
the horizon and flatness problems, Phys. Rev. {\bf D}23, 347
(1981); K. Sato, First-order phase transition of a vacuum
and the expansion of the universe, Mon. Not. R. Astron.
Soc. {\bf 195}, 467 (1981).

\bibitem{conformal}  R. Kallosh and A. Linde, Universality Class in Conformal
Inflation, JCAP {\bf 07}, 002 (2013) [arXiv:1306.5220 [hep-
th]]; R. Kallosh and A. Linde, Multi-field Conformal Cosmological Attractors, JCAP  {\bf 12}, 006 (2013)
[arXiv:1309.2015 [hep-th]].

\bibitem{alpha} D. I. Kaiser and E. I. Sfakianakis, Multifield Inflation after Planck: The Case for Nonminimal Couplings,
Phys. Rev. Lett.  {\bf 112}, no. 1, 011302 (2014) [arXiv:1304.0363
[astro-ph.CO]].

\bibitem{alpha1} S. Ferrara, R. Kallosh, A. Linde and M. Porrati, Minimal Supergravity Models of Inflation,  Phys. Rev. D{\bf 88}, no. 8,
085038 (2013) [arXiv:1307.7696 [hep-th]].

\bibitem{alpha2} R. Kallosh, A. Linde and D. Roest, Superconformal Inflationary $\alpha-$Attractors, JHEP  {\bf 11}, 198 (2013)
[arXiv:1311.0472 [hep-th]]. 

\bibitem{alpha3} R. Kallosh, A. Linde and D. Roest, Large field inflation and double $\alpha-$attractors, JHEP  {\bf 08}, 052 (2014) [arXiv:1405.3646 [hep-th]].

\bibitem{alpha4} T. Miranda, J. C. Fabris and O. F. Piattella,  {JCAP} {\bf 09} (2017) 041.

\bibitem{staro1980}A. A. Starobinsky, A new type of isotropic cosmological
models without singularity, Phys. Lett. B {\bf 91}, 99 (1980).

\bibitem{staro1} V. F. Mukhanov and G. V. Chibisov, {JETP Lett.} {\bf 33} (1981) 532.
 
\bibitem{staro2}  A. A. Starobinsky, { Sov. Astron. Lett.} {\bf 9} (1983) 302.
  
\bibitem{staro3} B. Whitt, {Phys. Lett. B} {\bf 145}  (1984) 176.
   
\bibitem{staro4} L. A. Kofman, A. D. Linde and A. A. Starobinsky, { Phys. Lett. B} {\bf 157} (1985) 361.

\bibitem{GL} A. S. Goncharov and A. D. Linde, Chaotic Inflation Of The
universe In Supergravity, Sov. Phys. JETP  {\bf 59}, 930 (1984)
[Zh. Eksp. Teor. Fiz. 86, 1594 (1984)]; A. B. Goncharov and
A. D. Linde, Chaotic Inflation in Supergravity, Phys. Lett.
B {\bf 139}, 27 (1984); A. Linde, Does the first chaotic inflation model in super-gravity provide the best fit to the Planck data?, JCAP  {\bf  02}, no. 02, 030 (2015) [arXiv:1412.7111 [hep-th]].

\bibitem{Planck2018} Planck Collaboration et al., Planck 2018 results. X. Constraints on
inflation, arXiv:1807.06211 [astro-ph].

\bibitem{alamPRD2018} M. Shahalam, M. Sami, A. Wang, Preinflationary dynamics of $\alpha-$attractor in loop quantum cosmology, Phys. Rev. D{\bf 98},  043524 (2018) [arXiv:1806.05815].

\bibitem{borde1994}A. Borde and A. Vilenkin, Eternal inflation and the initial singularity, Phys. Rev. Lett. {\bf 72}, 3305 (1994).

\bibitem{borde2003}A. Borde, A. H. Guth, and A. Vilenkin, Inflationary Spacetimes Are Incomplete in Past Directions, Phys. Rev. Lett. {\bf 90}, 151301 (2003).

\bibitem{martin2014}J. Martin, C. Ringeval, and V. Vennin, Encyclopaedia Inflationaris, Phys. Dark Univ. {\bf 5}  (2014) 75 [arXiv:1303.3787].

\bibitem{martin2001}J. Martin and R. H. Brandenberger, Trans-Planckian problem of inflationary cosmology, Phys. Rev. D{\bf 63}, 123501 (2001).

\bibitem{berger2013}R. H. Brandenberger and J. Martin, Trans-Planckian issues for inflationary cosmology, Class. Quantum Grav. {\bf 30}, 113001 (2013).

\bibitem{agullo2013a}I. Agullo, A. Ashtekar, and W. Nelson, Quantum Gravity Extension of the Inflationary Scenario, Phys. Rev. Lett.  {\bf 109}, 251301 (2012); Phys. Rev. D{\bf 87}, 043507
(2013).

\bibitem{agullo2013b}I. Agullo, A. Ashtekar, and W. Nelson, The pre-inflationary dynamics of loop quantum cosmology: confronting quantum gravity with observations, Class. Quantum Grav. {\bf 30}, 085014 (2013).

\bibitem{agullo2015}I. Agullo and N. A. Morris, Detailed analysis of the predictions of loop quantum cosmology for the primordial power spectra, Phys. Rev. D{\bf 92}, 124040 (2015).

\bibitem{ashtekar2011} A. Ashtekar and P. Singh, Loop quantum cosmology: a status report, Class. Quantum Grav. {\bf 28}, 213001 (2011).

\bibitem{ashtekar2015}A. Ashtekar and A. Barrau, Loop quantum cosmology:
from pre-inflationary dynamics to observations, Class. Quantum Grav. {\bf 32}, 234001 (2015).

\bibitem{barrau2016}A. Barrau and B. Bolliet, Some conceptual issues in loop
quantum cosmology, arXiv:1602.04452.


\bibitem{ashtekar2010}A. Ashtekar and D. Sloan, Loop quantum cosmology and
slow roll inflation, Phys. Lett. B  {\bf 694}, 108 (2010); Probability of inflation in loop quantum cosmology, Gen. Relativ. Gravit. {\bf 43}, 3619 (2011).

\bibitem{psingh2006}P. Singh, K. Vandersloot, and G. V. Vereshchagin, Non-singular bouncing universes in loop quantum cosmology, Phys. Rev. D{\bf 74}, 043510 (2006); 
J. Mielczarek, T. Cailleteau, J. Grain, and A. Barrau, Inflation in loop quantum cosmology: Dynamics and spectrum of gravitational waves, Phys. Rev. D{\bf 81}, 104049 (2010).

\bibitem{zhang2007} X. Zhang and Y. Ling, Inflationary universe in loop quantum cosmology, J. Cosmol. Astropart. Phys. {\bf 08}, 012 (2007).

\bibitem{chen2015} L. Chen and J.-Y. Zhu, Loop quantum cosmology: the horizon problem and the probability of inflation, Phys. Rev. D{\bf 92}, 084063 (2015) [arXiv:1510.03135 [gr-qc]].

\bibitem{bolliet2015}B. Bolliet, J. Grain, C. Stahl, L. Linsefors, and A. Barrau, Comparison of primordial tensor power spectra from the deformed algebra and dressed metric approachesin loop quantum cosmology, Phys. Rev. D{\bf 91}, 084035 (2015).

\bibitem{schander2016}S. Schander, A. Barrau, B. Bolliet, L. Linsefors, and J. Grain, Primordial scalar power spectrum from the Euclidean bounce of loop quantum cosmology, Phys. Rev.  D{\bf 93}, 023531 (2016).

\bibitem{bolliet2016}B. Bolliet, A. Barrau, J. Grain, and S. Schander, Observational Exclusion of a Consistent Quantum Cosmology Scenario, Phys. Rev. D{ {\bf 93}, 124011 (2016); J. Grain,
The perturbed universe in the deformed algebra approach of Loop Quantum Cosmology, Int. J. Mod. Phys. D  {\bf 25}, 1642003 (2016) [arXiv:1606.03271].

\bibitem{Bonga2016}B. Bonga and B. Gupt, Inflation with the Starobinsky
potential in Loop Quantum Cosmology, Gen. Relativ.
Gravit. {\bf 48}, 1 (2016); Phenomenological investigation of a
quantum gravity extension of inflation with the Starobin-
sky potential, Phys. Rev. D{\bf 93}, 063513 (2016).

\bibitem{Mielczareka}J. Mielczarek, Possible observational effects of loop quan-
tum cosmology, Phys. Rev. D{\bf 81}, 063503 (2010); L. Lin-
sefors, T. Cailleteau, A. Barrau, and J. Grain, Primor-
dial tensor power spectrum in holonomy corrected loop
quantum cosmology, Phys. Rev. D{\bf 87}, 107503 (2013); J.
Mielczarek, Gravitational waves from the big bounce, J.
Cosmol. Astropart. Phys. {\bf 11}, 011 (2008).


\bibitem{metrica}A. Ashtekar, W. Kaminski, and J. Lewandowski, Quan-
tum field theory on a cosmological, quantum space-time,
Phys. Rev. D{\bf 79} (2009) 064030.

\bibitem{metricb}I. Agullo, A. Ashtekar, and W. Nelson, Quantum Gravity
Extension of the Inflationary Scenario, Phys. Rev. Lett.
{\bf 109}, 251301 (2012).

\bibitem{metricc}I. Agullo, A. Ashtekar, and W. Nelson, Extension of
the quantum theory of cosmological perturbations to the
Planck era, Phys. Rev. D{\bf 87}, 043507 (2013).

\bibitem{algebraa}M. Bojowald, G. M. Hossain, M. Kagan, and S.
Shankaranarayanan, Gauge invariant cosmological per-
turbation equations with corrections from loop quantum
gravity, Phys. Rev. D{\bf 79}, 043505 (2009).

\bibitem{algebrab}J. Mielczarek, T. Cailleteau, A. Barrau, and J. Grain,
Anomaly-free vector perturbations with holonomy cor-
rections in loop quantum cosmology, Class. Quant. Grav.
{\bf 29}, 085009 (2012) [arXiv:1106.3744].

\bibitem{algebrac}T. Cailleteau, J. Mielczarek, A. Barrau, and J. Grain,
Anomaly-free scalar perturbations with holonomy cor-
rections in loop quantum cosmology, Class. Quant. Grav.
{\bf 29}, 095010 (2012) [arXiv:1111.3535].

\bibitem{algebrad}T. Cailleteau, A. Barrau, J. Grain and F. Vidotto, Con-
sistency of holonomy-corrected scalar, vector and tensor
perturbations in Loop Quantum Cosmology, Phys. Rev.
D{\bf 86}, 087301 (2012) [arXiv:1206.6736].

\bibitem{algebrae}T. Cailleteau, L. Linsefors, and A. Barrau, Anomaly-free
perturbations with inverse-volume and holonomy correc-
tions in loop quantum cosmology, Class. Quantum Grav.
{\bf 31}, 125011 (2014) [arXiv:1307.5238].

\bibitem{algebraf}A. Barrau, M. Bojowald, G. Calcagni, J. Grain, and M.
Khagan, Anomaly-free cosmological perturbations in ef-
fective canonical quantum gravity, J. Cosmol. Astropart.
Phys. {\bf 05} (2015) 051 [arXiv:1404.1018].

\bibitem{agullo15} I. Agullo, Loop quantum cosmology, non-Gaussianity, and CMB power asymmetry, 
Phys. Rev. D{\bf 92}, 064038 (2015)}.

\bibitem{ABS17} I. Agullo, B. Bolliet, and V. Sreenath, Non-Gaussianity in loop quantum cosmology, arXiv:1712.08148.


\bibitem{ZWKCS18} T. Zhu, A. Wang, K. Kirsten, G. Cleaver,  and Q. Sheng, Primordial non-Gaussianity and power asymmetry with quantum
gravitational effects in loop quantum cosmology, Phys. Rev. D{\bf 97}, 043501 (2018).

\bibitem{alam2017} M. Shahalam, M. Sharma, Q. Wu, A. Wang, Pre-inflationary dynamics in loop quantum cosmology: Power-law potentials, Phys. Rev. D{\bf 96},  123533 (2017) [arXiv:1710.09845]; M. Shahalam, Preinflationary dynamics of power-law potential in loop quantum cosmology, Universe 4 (2018) 87  [arXiv:1807.04620]. 

\bibitem{Tao2017a} T. Zhu, A. Wang, K. Kirsten, G. Cleaver, Q. Sheng, Universal features of quantum bounce in loop quantum cosmology, Phys. Lett. B{\bf 773} (2017) 196[arXiv:1607.06329]. 

\bibitem{Tao2017b} T. Zhu, A. Wang, G. Cleaver, K. Kirsten, Q. Sheng, Pre-inflationary universe in loop quantum cosmology, Phys. Rev. D{\bf 96}, 083520 (2017) [arXiv:1705.07544].

\bibitem{yang2009}J. Yang, Y. Ding, and Y. Ma, Alternative quantization of the Hamiltonian in loop quantum cosmology, Phys. Lett. B{\bf 682}, 1 (2009).

  
\bibitem{DL17} A. Dapor and K. Liegener, Cosmological Effective Hamiltonian from full Loop Quantum Gravity Dynamics, arXiv:1706.09833.
  
\bibitem{adlp} M. Assanioussi, A. Dapor, K. Liegener and T. Pawcowski, Emergent de Sitter epoch of the quantum Cosmos,   arXiv:1801.00768.

  
\bibitem{lsw2018a} B. F. Li, P.~Singh, A. Wang,  Towards cosmological dynamics from loop quantum gravity, Phys. Rev. D{\bf 97}, 084029 (2018) [arXiv:1801.07313].

\bibitem{lsw2018b} B. F. Li, P.~Singh, A. Wang,  Qualitative dynamics in pre-inflationary universe from loop quantum gravity, in preparation.
   
\bibitem{agullo18} I. Agullo, Primordial power spectrum from the Dapor-Liegener model of loop quantum cosmology, arXiv:1805.11356.


\bibitem{thiemann} T. Thiemann, Class. Quant. Grav. {\bf 15}, 839 (1998); T. Thiemann,
Class. Quant. Grav. {\bf 15}, 875 (1998); K. Giesel, T. Thiemann, Class. Quant. Grav. {\bf 24}, 2465 (2007).  

\bibitem{HISY}  K. Harigaya, M. Ibe, K. Schmitz, and T. T. Yanagida,
Chaotic inflation with a fractional power-law potential in
strongly coupled gauge theories, Phys. Lett. B{\bf 720}, 125
(2013); Dynamical fractional chaotic inflation, Phys. Rev. D{\bf 90}, 123524 (2014).

\bibitem{BG15} J. D. Barrow and A. A. H. Graham, Singular inflation, Phys. Rev.  D{\bf 91}, 083513 (2015); New singularities in unexpected places, Inter. J. M. Phys. D{\bf 24}, (2015) 1544012.

\bibitem{sahni18} S. S. Mishra, V. Sahni, A. V. Toporensky, Initial conditions for Inflation in an FRW universe [arXiv:1801.04948].

\bibitem{SW08} E. Silverstein and A. Westphal, Monodromy in the CMB: gravity waves and string inflation, Phys. Rev. D{\bf 78}, 106003 (2008).

\bibitem{killian} K. Martineau, A. Barrau and S. Schander, Phys. Rev. D{\bf 95}, 083507 (2017) [arXiv:1701.02703].
 	
\bibitem{nozari} K. Nozari, N. Rashidi, Perturbation, non-Gaussianity, and reheating in a Gauss-Bonnet $\alpha-$attractor model, Phys. Rev. D{\bf 95} (2017) 123518 [arXiv:1705.02617] 
[astro-ph.CO];  N. Rashidi, K. Nozari, $\alpha-$attractor and Reheating in a Model with Non-Canonical Scalar Fields, Int. J. Mod. Phys. D{\bf 27} (2018) 1850076 [arXiv:1802.09185] [astro-ph.CO].

\bibitem{BaoFei2019a} B. F. Li, P. Singh, A. Wang, Genericness of pre-inflationary dynamics and probability of the desired slow-roll inflation in modified loop quantum cosmologies; Phys. Rev. D100 (2019) no.6, 063513 [ arXiv:1906.01001].

\bibitem{BaoFei2019b} B. F. Li, T. Zhu, A. Wang, K. Kirsten, G. Cleaver, Q. Sheng, Preinflationary perturbations from the closed algebra approach in loop quantum cosmology; Phys. Rev. D99 (2019) no.10, 103536  [arXiv:1812.11191].

\bibitem{wu2018} Q. Wu, T. Zhu, A. Wang, Non-adiabatic evolution of primordial perturbations and non-Gaussinity in hybrid approach of loop quantum cosmology;  Phys. Rev. D98 (2018) no.10, 103528 [arXiv:1809.03172].

\bibitem{ma2019} W. J. Jin, Y. Ma, T. Zhu, Pre-inflationary dynamics of Starobinsky inflation and its generalization in Loop Quantum Brans-Dicke Cosmology; JCAP 1902 (2019) 010 [arXiv:1808.09643].


\bibitem{anshu2019} A. Bhardwaj, E. J. Copeland, J. Louko, Inflation in Loop Quantum Cosmology; Phys. Rev. D99 (2019) no.6, 063520 [arXiv:1812.06841].

\bibitem{Bea2018} B. E. Navascues, D. M. de Blas, G. A. M. Marugan, Time-dependent mass of cosmological perturbations in the hybrid and dressed metric approaches to loop quantum cosmology, Phys. Rev. D97 (2018) no.4, 043523 [arXiv:1711.10861];
B. E. Navascues, D. M. de Blas, G. A. M. Marugan, The Vacuum State of Primordial Fluctuations in Hybrid Loop Quantum Cosmology; Universe 4 (2018) no.10, 98
[arXiv:1809.09874].

\bibitem{sharma2018} M. Sharma, M. Shahalam, Q. Wu, A. Wang, Preinflationary dynamics in loop quantum cosmology: Monodromy Potential; JCAP 1811 (2018) 003
[arXiv:1808.05134].

\bibitem{ye2018} Y. Ye, T. Harko, Shi-Dong Liang, Loop quantum cosmology with a non-commutative quantum deformed photon gas; Eur. Phys. J. C78 (2018) no.7, 587
[arXiv:1807.05874].

\bibitem{kalloshPRL15}M. Galante, R. Kallosh, A. Linde, D. Roest, Phys. Rev. Lett. 114, 141302 (2015) [arXiv:1412.3797].

\bibitem{linder15} E. V. Linder, Dark Energy from $\alpha-$Attractor, Phys. Rev. D{\bf 91} (2015) 123012.

\bibitem{alam2018} M. Shahalam, R. Myrzakulov, S. Myrzakul, A. Wang, Observational constraints on the generalized $\alpha-$attractor model, Int. J. Mod. Phys. D{\bf 27}, (2018) 1850058.

\bibitem{varun2018} S. Bag, S. S. Mishra and V. Sahni, JCAP, 08, 009 (2018).

\bibitem{ashtekar2006} A. Ashtekar, T. Pawlowski, and P. Singh, Phys. Rev. D{\bf 74}, 084003 (2006).

\bibitem{Meissne}K. A. Meissne, Black hole entropy in loop quantum gravity, Class. Quantum Grav. {\bf 21}, 5245 (2004).

\bibitem{Domagala}M. Domagala, J. Lewandowski, Black hole entropy from quantum geometry, Class. Quantum Grav. {\bf 21}, 5233 (2004).

\bibitem{Mielczarek} J. Mielczarek, T. Cailleteau, J. Grain, and A. Barrau, Inflation in loop quantum cosmology: Dynamics and spectrum of gravitational waves, Phys. Rev. D{\bf 81}, 104049 (2010).



\end{thebibliography}
\end{document}